\documentclass[12pt]{article}
\pdfoutput=1

\usepackage{epsf, amssymb}
\usepackage{cite}
\usepackage{epsfig}
\usepackage{amsmath}
\usepackage{latexsym}
\usepackage[english]{babel}
\usepackage{graphicx,color}
\usepackage{url}

\newcommand{\arXiv}[1]{\href{http://www.arXiv.org/abs/#1}{#1}}
\usepackage[colorlinks=true, linkcolor=blue, bookmarks=true]{hyperref}

\makeatletter
\renewcommand\section{\@startsection {section}{1}{\z@}%
                               {-3.5ex \@plus -1ex \@minus -.2ex}
                               {2.3ex \@plus.2ex}%
                               {\normalfont\large\bfseries}}
\renewcommand\subsection{\@startsection{subsection}{2}{\z@}%
                                 {-3.25ex\@plus -1ex \@minus -.2ex}%
                                 {1.5ex \@plus .2ex}%
                                 {\normalfont\bfseries}}
\makeatother

\def\IZ{\relax\ifmmode\mathchoice
{\hbox{\cmss Z\kern-.4em Z}}{\hbox{\cmss Z\kern-.4em Z}}
{\lower.9pt\hbox{\cmsss Z\kern-.4em Z}} {\lower1.2pt\hbox{\cmsss
Z\kern-.4em Z}}\else{\cmss Z\kern-.4em Z}\fi}
\def\IR{\relax{\rm I\kern-.18em R}}

\def\one{{\hbox{ 1\kern-.8mm l}}}

\newlength{\bredde}
\def\slash#1{\settowidth{\bredde}{$#1$}\ifmmode\,\raisebox{.15ex}{/}
\hspace*{-\bredde} #1\else$\,\raisebox{.15ex}{/}\hspace*{-\bredde}
#1$\fi}

\newsavebox{\zzzbar}
\sbox{\zzzbar}
{\setlength{\unitlength}{0.9em}
\begin{picture}(0.6,0.7)
\thinlines
\put(0,0){\line(1,0){0.6}}
\put(0,0.75){\line(1,0){0.575}}
\multiput(0,0)(0.0125,0.025){30}{\rule{0.3pt}{0.3pt}}
\multiput(0.2,0)(0.0125,0.025){30}{\rule{0.3pt}{0.3pt}}
\put(0,0.75){\line(0,-1){0.15}}
\put(0.015,0.75){\line(0,-1){0.1}}
\put(0.03,0.75){\line(0,-1){0.075}}
\put(0.045,0.75){\line(0,-1){0.05}}
\put(0.05,0.75){\line(0,-1){0.025}}
\put(0.6,0){\line(0,1){0.15}}
\put(0.585,0){\line(0,1){0.1}}
\put(0.57,0){\line(0,1){0.075}}
\put(0.555,0){\line(0,1){0.05}}
\put(0.55,0){\line(0,1){0.025}}
\end{picture}}

\newcommand{\ena}{\end{eqnarray}}
\newcommand{\eq}[1]{(\ref{#1})}

\newcommand{\be}{\begin{equation}}
\newcommand{\ee}{\end{equation}}

\def\K{{\cal K}}

\def\H{{\cal H}}
\newcommand{\half}{\frac{1}{2}}
\newcommand{\Tr}{{\rm Tr}}

\def\be{\begin{equation}}
\def\ee{\end{equation}}
\def\a{\alpha}
\def\b{\beta}
\def\d{\delta}

\def\l{\lambda}

\def\r{\rho}

\def\vp{\varphi}

\def\O{{\cal O}}

\def\({\left (}
\def\){\right )}
\def\[{\left [}
\def\[{\right ]}

\def\ba{\begin{eqnarray}}
\def\ea{\end{eqnarray}}

\def\a{\alpha}
\def\b{\beta}

\def \B{{\bf B}}
\def \K{{\bf K}}

\def \r{{\bf r}}

\input amssym.def
\input amssym.tex

\newcommand{\bbibitem}[1]{\bibitem{#1}\marginpar{#1}}
\def\Bibitem#1{\bibitem{#1}%
  \smash{\hbox to0pt{\raise1ex\hbox{\tiny[#1]}\hss}}}

\def\Label#1{\label{#1}%
  \smash{\hbox to0pt{\raise1ex\hbox{\tiny[#1]}\hss}}}
\def\noLabels{\let\Label=\label}
\def\nobbibitem{\let\bbibitem=\bibitem}
 \def\noBibitem{\let\Bibitem=\bibitem}

\def\[{\left [}
\def\]{\right ]}
\def\({\left (}
\def\){\right )}
\def\del{\partial}
\def\l{\lambda}

\def\r{\rho}

\def\d{\hat{d}}

\def\r2{\sqrt{2}}

\def\del{\partial}

\def\Label#1{\label{#1}%
  \smash{\hbox to0pt{\raise1ex\hbox{\tiny[#1]}\hss}}}
\def\noLabels{\let\Label=\label}
\def\nobbibitem{\let\bbibitem=\bibitem}

\newcommand{\ads}[1]{{\rm AdS}_{#1}}

\newcommand{\bea}{\begin{eqnarray}}
\newcommand{\eea}{\end{eqnarray}}
\newcommand \nn {\nonumber}

\newcommand{\beq} {\begin{equation}}
\newcommand{\eeq} {\end{equation}}
\newcommand{\beqa} {\begin{eqnarray}}
\newcommand{\eeqa} {\end{eqnarray}}

\newcommand{\beqn}{\begin{eqnarray}}
\newcommand{\eeqn}{\end{eqnarray}}

\newcommand{\bal}{\begin{equation}\begin{aligned}}
\newcommand{\eal}{\end{aligned}\end{equation}}

\def\z{\chi}
\def\vr{\varrho}
\def\a{\alpha}
\def\b{\beta}

\def\d{\delta}

\def\eps{\epsilon}
\def\l{\lambda}

\def\vp{\varphi}

\def\r{\rho}

\def \B{{\bf B}}

\def\O{{\cal O}}

\def\O{{\cal O}}

\def\K{{\cal K}}
\def\H{{\cal H}}
\def\F{{\cal F}}
\def\B{{\cal B}}

\newcommand{\dsty}{\displaystyle}
\def\ads{\text{AdS}}

\begin{document}

\begin{titlepage}
\begin{flushright}
\end{flushright}
\vfill
\begin{center}
{\Large \bf Inhomogeneous holographic thermalization}

\vskip 10mm

{\large V.~Balasubramanian$^{a,b}$, A.~Bernamonti$^c$, J.~de Boer$^d$, B.~Craps$^e$,\\
\vspace{3mm}
L.~Franti$^{f,g}$, F.~Galli$^e$,  E.~Keski-Vakkuri$^{f,h}$, B.~M\"uller$^{i}$, A.~Sch\"afer$^j$ }

\vskip 7mm

$^a$ David Rittenhouse Laboratory, Univ. of Pennsylvania,
 Philadelphia, PA 19104, USA. \\
$^b$Laboratoire de Physique Th\'{e}orique, \'{E}cole Normale Sup\'{e}rieure, 75005 Paris, France.  \\
$^c$ Instituut voor Theoretische Fysica, KU Leuven,\\
Celestijnenlaan 200D, B-3001 Leuven, Belgium.\\
$^d$ Institute for Theoretical Physics, University of Amsterdam,
\\1090 GL Amsterdam, The Netherlands. \\
$^e$ Theoretische Natuurkunde, Vrije Universiteit Brussel, and \\
\hspace*{0.15cm}  International Solvay Institutes,
Pleinlaan 2, B-1050 Brussels, Belgium. \\
$^f$Department of Physics, P.O.Box 64, FIN-00014 University of Helsinki, Finland.\\
$^g$Helsinki Institute of Physics, P.O.Box 64, FIN-00014 University of Helsinki, Finland.\\
$^h$Department of Physics and Astronomy, Uppsala University, SE-75108 Uppsala, Sweden.\\
$^i$Department of Physics, Duke University, Durham, NC 27708-0305, USA.\\
$^j$Department of Physics, Regensburg University, 93040 Regensburg, Germany.\\

\vskip 3mm
\vskip 3mm
{\small\noindent  {\tt vijay@physics.upenn.edu, alice@itf.fys.kuleuven.be, J.deBoer@uva.nl, Ben.Craps@vub.ac.be, lasse.franti@helsinki.fi, fgalli@vub.ac.be, esko.keski-vakkuri@helsinki.fi, mueller@phy.duke.edu, andreas.schaefer@physik.uni-r.de}}

\end{center}
\vfill

\begin{center}
{\bf ABSTRACT}
\vspace{3mm}
\end{center}

The sudden injection of energy in a strongly coupled conformal field theory and its subsequent thermalization can be holographically modeled by a shell falling into anti-de Sitter space and forming a black brane. For a homogeneous shell, Bhattacharyya and Minwalla were able to study this process analytically using a weak field approximation. Motivated by event-by-event fluctuations in heavy ion collisions, we include inhomogeneities in this model, obtaining analytic results in a long wavelength expansion. In the early-time window in which our approximations can be trusted, the resulting evolution matches well with that of a simple free streaming model. Near the end of this time window, we find that the stress tensor approaches that of second-order viscous hydrodynamics. We comment on possible lessons for heavy ion phenomenology.
\end{titlepage}

\tableofcontents


\section{Introduction}
\label{intro}

The holographic gauge-gravity duality provides a framework to model the behavior of strongly coupled quantum liquids. Such liquids are studied in many
experimental settings: quark-gluon matter created in ultrarelativistic heavy ion collisions, strongly correlated electrons in metals, cuprates and
heavy-fermion materials, condensates of ultra-cold atoms. Therefore there is ample motivation for using holographic models to gain insight into
strongly coupled dynamics. Holographic models are known to have a very low shear viscosity to entropy density ratio in the strong coupling limit of the boundary quantum field theory \cite{Kovtun:2004de,Buchel:2003tz,Policastro:2001yc}, which is also close to the values reported for quark-gluon matter produced at the Relativistic Heavy Ion Collider (RHIC) and the Large Hadron Collider (LHC) \cite{Aronson}, for a cold atom gas near the unitarity limit \cite{Turlapov,TSchaefer,Cao}, and for the electronic fluid in graphene \cite{MMueller}. The study of transport coefficients has also evolved to a more direct relation between gravity and fluid dynamics: to the fluid/gravity correspondence \cite{Bhattacharyya:2008jc,Hubeny:2011hd} where one finds that a particular long wavelength limit of Einstein's equations turns them to hydrodynamic equations. One can even derive the complete second order hydrodynamic equations for conformal relativistic fluids \cite{Bhattacharyya:2008jc,Baier:2007ix} and generalize the standard M\"uller-Israel-Stewart theory. More recently attention has been given to equilibration and thermalization from far-from-equilibrium initial conditions. A particularly interesting question is how and when a strongly coupled quantum liquid approaches a regime in which hydrodynamics becomes a good approximation. This paper investigates that question in a particular analytically tractable model, with additional motivation coming from the early dynamics of hot quark-gluon matter in heavy ion collisions.

Relativistic heavy ion collisions have the dual advantage of forming almost perfectly isolated microscopic quantum systems, for which the von Neumann entropy should be exactly conserved, and of being studied in great detail experimentally at RHIC and at the LHC. Experiments at these accelerators have shown that already at very early times, at most 1 fm/c, the matter produced in the heavy ion collisions shows collective behavior in agreement with what is expected from viscous hydrodynamics. The validity of a hydrodynamical description generally relies on the fluid being near local thermal equilibrium, but some degree of collectivity may exist even before local thermal equilibrium is reached. This conjecture is supported by recent experiments studying p+Pb collisions at LHC \cite{Aad:2012gla,Abelev:2012ola} and d+Au collisions at RHIC \cite{Adare:2013piz}, which show indications of the presence of hydrodynamic behavior in events producing a large particle multiplicity. On the other hand, several theoretical approaches to the dynamics of thermalization in QCD, e.g.\ the perturbative bottom-up thermalization framework \cite{Baier:2002bt} and classical gauge theory  \cite{Kunihiro:2010tg}, predict substantially longer equilibration times. Heavy ion experiments thus offer a playground for the study of equilibration and thermalization in gauge theories.

The argument about the presence of hydrodynamic behavior of the quark-gluon plasma created in heavy ion collisions rests primarily on two observations. First, one observes a $\cos(2\phi)$ correlation between the azimuthal momentum direction of produced hadrons and the collision plane, which is known as ``elliptic'' flow, see \cite{Adare:2006ti,Adamczyk:2013gw,ALICE:2011ab,ATLAS:2012at} and references therein. This phenomenon can also be deduced from the azimuthal two-particle correlations among emitted hadrons, and is sometimes referred to as the ``ridge'' or ``double-ridge'' effect in studies in Pb+Pb collisions at  LHC \cite{ATLAS:2012at,Chatrchyan:2012wg,Abelev:2009af,Aamodt:2011by}. The second observation is related to event-by-event fluctuations. Experimentally it was found for symmetric heavy ion collisions that odd Fourier coefficients of the flow are not much smaller than even ones \cite{ATLAS:2012at,Abelev:2012di,Chatrchyan:2012wg}. By a symmetry argument, odd coefficients can only be generated by fluctuations, so one is forced to conclude that fluctuations are large. The parton saturation model for the initial nuclear state suggests also that they are of short range in the plane transverse to the beam axis (of order of the inverse saturation scale $1/Q_s$).  Detailed simulations \cite{Schenke:2012hg,Gale:2012rq,Bzdak:2013zma,Heinz:2013th,Adare:2012kf} show that if one assumes also these small size fluctuations to evolve hydrodynamically one gets excellent fits to all flow coefficents.

The holographic description of strongly coupled gauge theories offers a framework for the study of the rate of thermalization and the approach to hydrodynamical flow.  In \cite{AbajoArrastia:2010yt, Albash:2010mv, Balasubramanian:2010ce, Balasubramanian:2011ur}, it was found that in a simple model thermalization after the homogeneous, isotropic injection of energy  proceeds very fast. The initial state of heavy ion collisions is, however, quite anisotropic and inhomogeneous. It is characterized by a strong asymmetry between longitudinal and transverse pressure (the former may even be negative initially due to the presence of strong longitudinal gauge fields) and by large density fluctuations in the transverse direction. The effect of both can be studied with a refined AdS/CFT treatment.  The effect of the pressure anisotropy was studied in \cite{Chesler:2009cy,Beuf:2009cx,Heller:2011ju} under the assumption of longitudinal boost invariance and transverse homogeneity; it was found that hydrodynamic behavior is reached on time scales of order $0.3-0.5$ fm/c for many different initial conditions.  This ``hydroization'' is not equivalent to complete thermalization because viscous hydrodynamic behavior at early times in a boost invariant expansion implies a rather large pressure anisotropy and thus strong deviation from local thermal equilibrium.

In the present paper we analyze the second aspect, namely the question how the approach to hydrodynamic behavior is affected by local density fluctuations, which has not been studied in detail so far. A priori, there is no reason to believe that hydrodynamics, which is  a low-energy effective description, should describe the short-time evolution we will be studying. However, since hydrodynamics turned out to apply unexpectedly early in homogeneous models, it is interesting to ask to what extent this surprise extends to inhomogeneous setups.

\begin{figure}[ht]
\centering
\includegraphics[width= 0.4\textwidth]{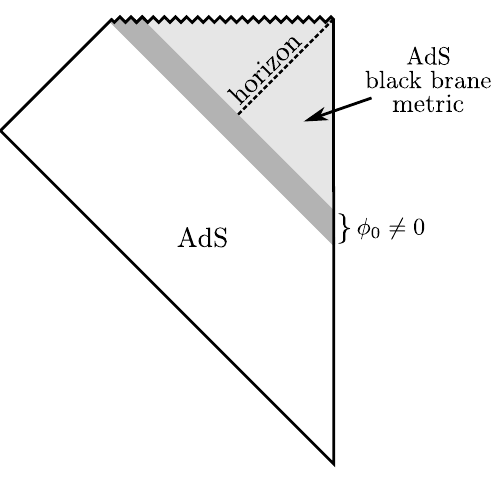}
\caption{A schematic representation of the dynamical collapse process first studied in \cite{Bhattacharyya:2009uu}. In pure AdS spacetime, a scalar source $\phi_0$ on the boundary, corresponding to the asymptotic boundary value of a massless bulk scalar field, is turned on for a lapse of time $\d t$. This induces a non-vanishing profile for the bulk scalar field  which backreacts on the AdS geometry as a shock wave propagating in the bulk leading to the formation of a black hole horizon.}
\label{fig:collapse}
\end{figure}

Thermalization in strongly coupled conformal field theories with a gravity dual corresponds to black brane formation in asymptotically anti-de Sitter (AdS) spacetimes. To study gravitational collapse, one generically needs numerical general relativity, but interesting situations exist where analytic computations are possible. Consider a massless scalar  minimally coupled to gravity in $d+1$ spacetime dimensions with negative cosmological constant. In this setting, the authors of \cite{Bhattacharyya:2009uu} considered the effect of a homogeneous boundary source on the bulk geometry of an asymptotically AdS spacetime. Specifically, they turned on for a finite lapse of time $\d t$ a homogeneous source $\phi_0(t)$ for a marginal boundary operator corresponding to the massless scalar field in the bulk. Solving the field equations in a weak field approximation, {\it i.e.} in the limit where the amplitude of the scalar source is small in an appropriate sense\footnote{The ``injection time'' $\d t$ should be small compared to the inverse temperature of the black brane that will eventually be formed.},  in \cite{Bhattacharyya:2009uu} it  was shown that this generically leads to black brane formation in the bulk (see also \cite{Garfinkle:2011hm, Garfinkle:2011tc, Wu:2012rib} for numerical analyses). For $d+1=4$ and to leading order in the amplitude of the scalar source, the geometry is given by the AdS$_4$-Vaidya metric
\be\label{AdSVaidya}
ds^2 =  - \(r^2 - \frac{M(v)}{ r} \) dv^2+ 2 dv dr + r^2 ( dx^2  + dy^2)\, ,
\ee
with $M(v)$ a non-decreasing function that vanishes for $v \le 0$. This geometry describes a shell of null dust falling in from the boundary of AdS and collapsing into a black brane. A schematic representation of the process is depicted in Fig.~\ref{fig:collapse}.

The AdS-Vaidya model has been very useful as a simple, tractable model of holographic thermalization. Many observables in the dual field theory have been identified whose time-evolution can be computed relatively easily from geometric quantities in AdS-Vaidya \cite{Hubeny:2007xt, AbajoArrastia:2010yt, Albash:2010mv, Balasubramanian:2010ce, Balasubramanian:2011ur, Balasubramanian:2011at, Allais:2011ys, Aparicio:2011zy, Keranen:2012tv, Callan:2012ip, Arefeva:2012jp, Balasubramanian:2012tu, Hubeny:2013hz,Liu:2013iza}. When this model is used as a very crude analogue of the equilibrating quark-gluon plasma in heavy ion collisions, an encouraging result is that, at least for the range of length scales studied in \cite{ Balasubramanian:2010ce, Balasubramanian:2011ur}, thermalization for homogeneous energy density occurs as fast as allowed by causality.\footnote{For $d>2$, it was found in \cite{Liu:2013iza} that on length scales larger than the inverse temperature, thermalization happens at a smaller speed.} This insight suggests that hydrodynamics can already describe early stages of heavy ion collisions with strong coupling dynamics. Massive infalling shells were studied in \cite{Danielsson:1999zt, Danielsson:1999fa, Lin:2008rw, Erdmenger:2012xu, Steineder:2012si, Baier:2012ax, Baier:2012tc, Baron:2012fv, Steineder:2013ana}, charged shells in \cite{Caceres:2012em, Galante:2012pv, Caceres:2012px, Lin:2013sga} and shells with angular momentum in \cite{Aref'eva:2013wma}.

Several holographic thermalization models have been proposed that are probably more realistic than the AdS-Vaidya model. Examples include anisotropic homogeneous plasmas \cite{Janik:2008tc, Chesler:2008hg, Chesler:2011ds, Chesler:2012zk, Heller:2012km, vanderSchee:2012qj, Heller:2013oxa}, boost invariant plasmas \cite{Janik:2006gp, Benincasa:2007tp, Chesler:2009cy, Beuf:2009cx, Heller:2011ju, Heller:2012je}, and colliding shock waves \cite{Albacete:2008vs, Lin:2009pn,Albacete:2009ji,Taliotis:2010pi, Chesler:2010bi, Kiritsis:2011yn, Romatschke:2013re, Taliotis:2012sx,Casalderrey-Solana:2013aba}. Still, a high degree of symmetry is usually assumed to keep the (mostly numerical) computations tractable. Also in these models, fast thermalization is often found \cite{Chesler:2009cy, Chesler:2010bi, Heller:2011ju, Chesler:2011ds}. Holographic thermalization has also been used to model condensed matter systems \cite{Keranen:2011xs, Bhaseen:2012gg, Basu:2011ft, Basu:2012gg, Li:2013fhw}. Other models related to holographic thermalization are studied in \cite{Das:2010yw, Ebrahim:2010ra, Erdmenger:2011jb, AliAkbari:2012hb, Buchel:2013lla, Buchel:2012gw, Nozaki:2013wia, Nozaki:2013vta, Zeng:2013mca, Li:2013sia,Shenker:2013pqa, Brady:2013opa, Hung:2013dka, Fukushima:2013dma}.

In this paper we generalize  the construction of \cite{Bhattacharyya:2009uu} to the case of an inhomogeneous scalar field source at the boundary. As in \cite{Bhattacharyya:2009uu}, we solve the equations of motion in a perturbative expansion in the amplitude of the scalar field boundary value, which is here allowed to depend on the spatial coordinates. Before the source is turned on, the solution coincides with the pure AdS background. When introducing a dependence on the transverse spatial coordinates, the situation becomes much more involved and finding an exact analytic solution is complicated. In order to still get an analytic insight, in this paper we shall therefore attack the problem under the simplifying assumption that the scale of spatial variations is large as compared to all other scales, following a strategy suggested in \cite{Bhattacharyya:2009uu}.

As in \cite{Bhattacharyya:2009uu}, the case of a four-dimensional bulk space-time turns out to be technically simpler than the five-dimensional one.\footnote{Since in heavy ion collisions azimuthal anisotropies are studied in the directions transverse to the beam, it is not even obvious that for our purposes a four-dimensional bulk geometry should be less relevant than a five-dimensional geometry without a beam direction singled out. More realistic models would involve nearly boost invariant setups or colliding shock waves, but for computational tractability these will not be studied in the present paper.} Ideally, one would like to introduce space dependence for at least two directions, corresponding to the plane in which the pancaked nuclei overlap in the collision, but for simplicity we consider the simplest case involving spatial dependence on a single coordinate (in addition to the radial coordinate in the bulk). So we consider an asymptotically AdS$_{4}$ geometry with inhomogeneities along a single spatial direction.

In section~\ref{sec:gradient}, we construct the bulk solution up to second order in the amplitude of the source that drives the gravitational collapse and up to fourth order in the gradient expansion. We argue that our solutions should be reliable for times short compared to the inverse local temperature (of the black brane to be formed) and wavelengths large compared to the inverse local temperature.
In section~\ref{sec:stress}, we extract from the bulk solution the expectation value of the boundary stress-energy tensor, and study its time-evolution after the inhomogeneous energy injection. We compare this evolution to that of a simple free-streaming model as well as to first and second order hydrodynamics in section~\ref{sec:comparison}. Section~\ref{sec:summary} contains a summary and a discussion of the possible relevance for heavy ion phenomenology.

A short account of our main results can be found in the companion paper\cite{Balasubramanian:2013rva}.


\setcounter{equation}{0}


\section{AdS$_4$ weak field inhomogeneous collapse in the gradient expansion} \label{sec:gradient}

We consider a massless scalar  minimally coupled to gravity in four spacetime dimensions with negative cosmological constant,
\be \label{bmact}
S = \frac{1}{16 \pi G_{N}}\int d^{4}x \sqrt{g}\left(R  - 2 \Lambda-\frac{1}{2}g^{\mu\nu}\del_{\mu}\phi \del_{\nu}\phi \right)\, .
\ee
Here $\Lambda = -3$ and the AdS radius has been set to one. The equations of motion following from \eqref{bmact} read
\bal \label{eminh}
E_{\mu\nu}&\equiv G_{\mu\nu} -\frac{1}{2}\del_{\mu}\phi \del_{\nu}\phi + g_{\mu\nu}\left(- 3+\frac{1}{4}(\del\phi)^{2}\right)=0\, ,
 \\
\Box\phi& = \frac{1}{\sqrt{g}}\del_{\mu}(\sqrt{g} g^{\mu \nu} \del_{\nu}\phi )=0 \,.
\eal

When assuming inhomogeneities along only one of the two boundary spatial directions, say $x$, the ansatz in Eddington-Finkelstein coordinates consistent with the symmetries of the problem can be written as
\bal  \label{EFgr}
&ds^{2} =   -h(v, r,x)dv^{2} +   2dv \( dr + k(v, r,x) dx\)  +  \\[5 pt]
&\hspace{2.5 cm}+f(v,r,x)^2 e^{B(v,r,x)}dx^2+f(v,r,x)^2 e^{-B(v,r,x)}dy^2  \, , \\[5 pt]
&\phi = \phi(v,r,x)\, ,
\eal
where $v$ is an ingoing null coordinate, $r$ the AdS radial coordinate and $y$ denotes the other spatial direction.

On the boundary, where the null coordinate $v$  is identified with the gauge theory time $t$, we turn on an inhomogeneous scalar field source
\be
\label{scalarsource}
\begin{aligned}
&\vp(v,x)=0 \, ,\\
&\vp(v,x)=\eps \varphi_{0}(v,x)\, ,   \\
&\vp(v,x)=0 \, ,
\end{aligned}
\qquad
\begin{aligned}
& v \le 0 \\
&  0<v< \delta t  \\
&   v \ge \delta t  \, ,
\end{aligned}
\ee
over a time $ \delta t $.
Here we have singled out  the explicit factor  $\eps$ parametrizing the amplitude of the scalar source for the bulk field,  which with the non-canonical normalization \eqref{bmact} of the bulk kinetic term is dimensionless. As in \cite{Bhattacharyya:2009uu}, we require the forcing function $\varphi_{0}(v,x)$ and its first few time derivatives to be everywhere continuous.

The  initial  condition that the spacetime should be pure AdS  for $v \le 0$ corresponds  to
\be \label{geninitial}
\begin{aligned}
& h(v \le 0, r,x) =r^{2} \\
& k(v\le 0, r,x)=0
\end{aligned} \qquad \qquad
\begin{aligned}
& f(v\le 0,r,x) = r  \\
& B(v\le0,r,x)  =  0  \, ,
\end{aligned}
\ee
and
\be \label{geninitialphi}
\phi(v\le0, r, x) = 0 \,,
\ee
while the  asymptotically AdS  boundary conditions (with planar boundary geometry) read
\be \label{genasympt}
\begin{aligned}
&\lim_{r \to \infty}\frac{ h(v, r,x)}{r^2} =1\\
&\lim_{r \to \infty}\frac{ k(v, r,x)}{r^2} =0
\end{aligned}  \qquad\qquad
\begin{aligned}
&\lim_{r \to \infty}\frac{ f(v, r,x)^{2}}{r^2} = 1 \\
&\lim_{r \to \infty} B(v, r,x) = 0 \, ,
\end{aligned}
\ee
and
\be  \label{genasymptphi}
\lim_{r \to \infty} \phi(v, r, x ) = \vp(v,x) \,.
\ee
The  metric \eqref{EFgr} supplemented with the boundary conditions \eqref{genasympt} is not completely gauge fixed.  The form of the metric is left unchanged under the transformation $r \to r + s(v,x)$.  We use this residual freedom to choose the subleading behavior of $ f(v,r,x)$ to be  $f(v,r,x) = r(1+ O(1/r^2))$.

With an ansatz  of  the form \eqref{EFgr} one has, in addition to the scalar equation of motion, seven  equations coming from \eqref{eminh}.  In fact, of the ten components  of  $E_{\mu\nu}$, three turn out to identically vanish and one of the remaining is a linear combination of the others together with the scalar equation. The following linear combinations turn out to be a  convenient choice to work with:
\be   \label{einstindep}
\begin{aligned}
&E_{\phi} \equiv \Box\phi =0  \\
&E_{1} \equiv g^{v\mu}E_{\mu r} =0  \\
&E_{2} \equiv g^{v\mu}E_{\mu v } =0 \\
&E_{3} \equiv g^{v\mu}E_{\mu x} =0
\end{aligned}
\qquad
\qquad
\begin{aligned}
&E_{4} \equiv g^{x\mu}E_{\mu x } =0 \\
&E_{5} \equiv g^{y\mu}E_{\mu y } =0 \\
&E_{c1} \equiv g^{r \mu}E_{\mu v }  =0 \\
&E_{c2} \equiv g^{x \mu}E_{\mu v} =0 \, ,
\end{aligned}
\ee
where $E_{ci}=0$, $i=1,2$ are conservation equations, and one of the equations is implied by the others.

The approach we adopt here is to solve the system  \eqref{einstindep} in a double expansion. Following \cite{Bhattacharyya:2009uu}, we perform a weak field expansion in the amplitude $\epsilon$ of the scalar field source. On top of this, we consider a spatial gradient expansion along the non-homogeneous space direction, {\it i.e.} an expansion in spatial derivatives\footnote{The combination of amplitude and gradient expansions has also been used in \cite{Mukhopadhyay:2012hv,Banerjee:2012uq} to study non-equilibrium Green functions and response.}. The metric components and scalar field are then written in a double expansion in the parameters $\epsilon$, which keeps track of the order in the amplitude of the source, and $\mu$, which acts as a formal derivative counting parameter, as
\be \label{doubleexp}
\begin{aligned}
h(v, r,x) &=& \sum_{n,i=0}^{\infty}  \eps^{n} \mu^{i} h_{n,i}(v, r, \mu x) \\
f(v, r,x) &=& \sum_{n,i=0}^{\infty}  \eps^{n} \mu^{i} f_{n,i}(v, r, \mu x)  \\
\phi(v, r,x) &=& \sum_{n,i=0}^{\infty} \eps^{n} \mu^{i} \phi_{n,i}(v, r, \mu x)
\end{aligned} \qquad
\begin{aligned}
k(v, r,x) &=& \sum_{n,i=0}^{\infty} \eps^{n} \mu^{i} k_{n,i}(v, r, \mu x)  \\
B(v, r,x) &=& \sum_{n,i=0}^{\infty}  \eps^{n} \mu^{i} b_{n,i}(v, r, \mu x) \, .
\end{aligned}
\ee
Accordingly, the boundary source is written as
\be
\begin{aligned}
&\vp(v,\mu x)=0 \, ,\\
&\vp(v,\mu x)=\eps \varphi_{0}(v,\mu x)\, ,   \\
&\vp(v,\mu x)=0 \, ,
\end{aligned}
\qquad
\begin{aligned}
& v \le 0 \\
&  0<v< \delta t  \\
&  v \ge \delta t \, ,
\end{aligned}
\ee
and we assume that the function $\varphi_{0}(v,\mu x)$ and its derivatives are everywhere continuous.
The derivative counting parameter $\mu$ shall be set to one in the solution at the end of the computations. For compactness in what follows we refer to the order in the number of derivatives as the order in $\mu$.


\subsection{General structure of the equations and of the solution}

The different coefficients in the expansion \eqref{doubleexp} are determined order by order as a function of the previous orders solution.  Implementing the pure AdS initial conditions  \eqref{geninitial} and \eqref{geninitialphi}, as well as the asymptotically AdS conditions \eqref{genasympt} and \eqref{genasymptphi},  this implies that these coefficients  involve only   the forcing function $\vp$ for the bulk scalar profile.

There are a few general considerations about the structure of the two expansions \eqref{doubleexp} worth making  before explicitly solving  the equations.
The background  solution of the derivative expansion ($O(\mu^0)$)  coincides with the homogeneous solution of  \cite{Bhattacharyya:2009uu},  with the  $x$-dependence added by hand to the source $\vp$. At zeroth order in $\mu$ and first order in $\eps$,  $\phi_{1,0}$ is forced by its boundary condition to be non-zero.  The background geometry is pure AdS at this order, since there is no $O(\eps)$ source for the Einstein's tensor coming from the bulk scalar.  Only at order $O(\eps^{2})$ are the metric components  $h_{2,0}$ and $f_{2,0}$ sourced.

When we add the $x$-dependence to the homogeneous solution, it ceases to be a solution to the equation of motion.  At leading order in $\eps$ the equation for $\phi$ is just the  massless scalar equation in pure AdS${}_{4}$. By symmetry of the background, the terms violating the homogeneous scalar equation are of order $\mu^2$, {\it i.e.}\  second $x$-derivatives of the source. The background bulk scalar solution thus needs to be corrected with a term of order $\mu^2$.  A similar reasoning applies recursively at higher orders and for the metric which starts at order $\eps^2$. Notice that the scalar profile at order $\eps$ receives only even order corrections in $\mu$. Similarly all metric coefficients at order $\eps^2$ receive only even order corrections in $\mu$, with the exception of $k(v,r,x)$, which has only odd $\mu$ contributions.

Solving  the equations \eqref{einstindep}, one recognizes a structure similar to that observed in \cite{Bhattacharyya:2009uu}. In the amplitude and gradient expansion, the scalar equation of motion is  solved  using a  $1/r$ expansion and this automatically leads to metric components  expressed as an expansion in $1/r$. As illustrated in  Appendix \ref{app:bulk}, the initial conditions are satisfied by imposing that the $1/r$ expansions should terminate at a finite order.

Order by order in $\eps$ and $\mu$,  $E_{1}=0$ gives a differential equation in $r$ for the metric component $f_{n,i}$, which is determined completely by imposing the asymptotic boundary conditions.  $E_{2}=0$ similarly gives a differential equation in $r$ for $h_{n,i}$, which is completely fixed imposing the boundary conditions together with  the conservation equation $E_{c1}=0$. In the same way, the coefficients  $k_{n,i}$ are determined using $E_{3}=0$ and $E_{c2}=0$. The former gives a differential equation in $r$ for $k_{n,i}$ while the latter is a conservation equation, which together with the boundary conditions  completely fixes the form of $k_{n,i}$.  Using the rest of the equations,  $E_{4}=0$ and $E_{5}=0$ follow from each other. The resulting independent equation is the only equation involving $b_{n,i}$ and suffices to determine its expression.


\subsection{Solution at first and second order in the source amplitude} \label{subsec:solution}

Solving the system  \eqref{einstindep}  following the strategy outlined  in the previous Section, we obtain the following solution up to second order in  the amplitude of the source field and up to fourth order in the spatial derivative expansion.    The compact form of  the solution with the different orders in the $\eps$ and $\mu$ expansions grouped according to their radial asymptotic behaviour is given in Appendix \ref{app:bulk} for convenience. Here we present the solution as obtained order by order in the two expansion  parameters.

In the following we adopt the  notation $\del_{v} F(v,x)  =  \dot{F}(v,x)$  and $\del_{x}F(v,x) = F'(v,x)$.

\subsection*{First order in $\eps$ }

At zeroth order in  $\mu$ (we set $\mu =1$ in writing the solution)
 \be \label{phi10}
\eps \phi_{1,0}(v,r, x) = \vp(v, x)+ \frac{\dot \vp(v, x)}{r}\, ,
\ee
which, when removing the $x$-dependence, coincides with the homogeneous solution obtained in \cite{Bhattacharyya:2009uu}.
The  next non-vanishing contribution is at  $O\(\mu^2\)$
\be \label{phi12}
\eps \phi_{1,2}(v,r,  x) =  \frac{\vp''(v, x)}{2r^2}\, ,
\ee
while at fourth order in $\mu$ one obtains the additional contribution
\be \label{phi14}
\eps \phi_{1,4}(v,r,  x) =  \frac{\int_{-\infty}^{v} d\tau ~ \vp''''(\tau, x)}{ 8 r^3}\, .
\ee
All the metric coefficients vanish at this order in $\eps$.

\subsection*{Second order in $\eps$ }

The scalar profile in the bulk does not  get contributions at this order, while the metric at lowest (zeroth) order in $\mu$  is the one of the homogeneous case, with the spatial dependence added. Explicitly, the solution of \eqref{einstindep} reads
\bal \label{sol22}
\eps^2 f_{2,0}(v,r, x) &= -\frac{\( \dot\vp(v,x)\)^2}{ 8r} \, ,  \\
 \eps^2 h_{2,0}(v,r,x) &= -\frac{3\(\dot \vp(v,x)\)^2}{4} + \frac{C_{2,0}(v,x)}{r}\, ,
\eal
where
\be \label{C20}
  C_{2,0}(v,x) = \frac{1}{2} \int_{-\infty}^{v} d\tau~\dot\vp(\tau,x)\dddot \vp(\tau,x)\, ,
\ee
and all other coefficients vanish.

The coefficient of the  cross component $dvdx$ of the metric gets the first  non-vanishing contribution at order $\mu$
\be
 \eps^2  k_{2,1}(v,r, x) = \frac{1}{2}\dot\vp(v,x)\vp'(v,x) + \frac{K_{2,1}(v,x)}{6 r}\, ,
\ee
where
\bal
 K_{2,1}(v,x) &= \int_{- \infty}^{v} d\tau~ \left[-2 C_{2,0}'(\tau, x) + \dot\vp'(\tau,x)\ddot\vp(\tau, x) - \dot\vp(\tau,x)\ddot\vp'(\tau, x)\right] \\
&= \int_{- \infty}^{v} d\tau~ 2\left[    - \dot\vp(\tau,x)\ddot\vp'(\tau, x) + \int_{- \infty}^{\tau} dt ~\ddot\vp'(t,x)\ddot\vp(t, x) \right]\,  .
\eal
At order $\mu^2$ all other coefficients get sourced in Einstein's equations. Their expressions are
\bal
\eps^2  f_{2,2}(v,r, x) &= -\frac{\dot\vp(v,x) \vp''(v,x)}{12 r^2} \, , \\
\eps^2  h_{2,2}(v,r,x) &= -\frac{\(\vp'(v, x)\)^{2}}{4} +\frac{C_{2,2}(v,x)}{r}\\
&~~~~ -\frac{\dot \vp(v,x)\dot\vp''(v,x)- 2\ddot \vp(v,x) \vp''(v,x)}{12 r^2} -\frac{ K_{2,1}'(v, x) }{12 r^2} \, , \\
 \eps^2  b_{2,2}(v,r,x) &= \frac{ \( \vp'(v,x)\)^{2} }{4r^2} + \frac{\beta_{2,2}(v, x)}{ r^3}\, ,
\eal
with
\bal \label{C22}
 C_{2,2}(v,x) &= \frac{1}{4} \int_{-\infty}^{v} d\tau~\left[ -\( \dot\vp'(\tau, x) \)^2    -  4 \dot\vp (\tau, x) \dot\vp'' (\tau, x) \right.\\[3 pt]
& \hspace{2.5 cm } + 2 \ddot\vp (\tau, x)\vp''(\tau, x) + 2\ddot\vp'(\tau, x)\vp'(\tau, x) -K_{2,1}' (\tau, x) \Big]  \, ,  \\
 \beta_{2,2}(v,x)&= \frac{1}{8}\int_{-\infty}^{v} d\tau~ \left[ \(\dot\vp'(\tau,x) \)^2 +\frac{1}{3} K_{2,1}'(\tau, x)\right] \, .
\eal
Going further with the expansion in spatial derivatives, a third order contribution to the coefficient for the cross component $dvdx$ of the metric comes into play
\be
\eps^2  k_{2,3}(v,r, x) =  \frac{K_{2,3}(v,x)}{6 r}   - \frac{ \dot\vp'(v,x) \vp''(v,x) }{ 8 r^2 } + \frac{3 \beta_{2,2}(v, x)}{ 4 r^2} \, ,
\ee
where
\bal
 &K_{2,3}(v,x) = \int_{- \infty}^{v} d\tau~ \Big[-2 C_{2,2}'(\tau, x) + 6  \beta_{2,2}'(v,x) \\
 &\hspace{4 cm } +  \dot\vp(\tau,x)\vp'''(\tau, x) -2 \dot\vp'(\tau,x) \vp''(\tau, x)\Big]  \, .
\eal
Finally, at fourth order in the gradient expansion
\bal
\eps^2  f_{2,4}(v,r, x) &= -\frac{\(\vp''(v,x) \)^2}{48 r^3} -\frac{\dot \vp(v,x)  \int_{- \infty}^{v} d\tau~ \vp''''(v,x)}{64 r^3} \, , \\
\eps^2  h_{2,4}(v,r,x) &= \frac{C_{2,4}(v,x)}{r} -\frac{K_{2,3}'(v,x)}{2r^2} -\frac{\(\vp''(v,x)\)^2}{6 r^2}    \\
&~~~~  + \frac{  \dot \vp(v,x)  \int_{- \infty}^{v} d\tau~  \vp''''(t,x)  }{16 r^2}   -\frac{\beta_{2,2}''(v,x)}{4 r^3 }  + \frac{\(\dot \vp' (v,x)\vp'' (v,x)\)' }{24 r^3} \\
& ~~~~ - \frac{ \vp''''(v,x) \dot \vp(v,x)   + }{96 r^3} + \frac{  \ddot \vp(v,x)  \int_{- \infty}^{v} d\tau~  \vp''''(t,x)  }{32 r^3}  \, ,   \\
 \eps^2  b_{2,4}(v,r,x) &= \frac{ \beta_{2,4,3}(v,x) }{r^3} + \frac{\beta_{2,4,4}(v,x)}{ r^4}\, ,
\eal
with
\bal \label{C24}
C_{2,4}(v,x) &= \frac{1}{4} \int_{-\infty}^{v} d\tau~\Big[ -\( \vp''(\tau,x) \)^2    -   \vp' (\tau,x)\vp''' (\tau,x)  -  6 K_{2,3}' (\tau,x)  \\
&\hspace{5.5 cm } +  \frac{3}{4} \dot\vp(\tau,x) \int_{-\infty}^{\tau} dw~ \vp''''(w,x)   \Big]  \, , \\
 \beta_{2,4,3}(v,x)&= \frac{1}{8} \int_{-\infty}^{v} d\tau~ \left[ \(\vp'(\tau,x)\vp'''(\tau,x) \) + 2 K_{2,3}'(\tau, x) +  8  \beta_{2,4,4}(\tau,x)\right]  \, , \\
 \beta_{2,4,4}(v,x)&= \frac{1}{24} \int_{-\infty}^{v} d\tau~ \left[ \(\dot \vp'(\tau,x)\vp'''(\tau,x) -\ \dot\vp''(\tau,x) \vp''(\tau,x) \)  + 6  \beta_{2,2}''(\tau,x)\right] \, .
\eal
A simple consistency check on the different terms entering in the components of  the metric  and in the bulk scalar comes from  their scaling, with derivatives contributing with weight 1 and factors of $1/r$ and integrations contributing with weight $-1$.

All solutions reduce exactly to the homogeneous formulae \cite{Bhattacharyya:2009uu} when the spatial dependence of the scalar source is suppressed.


\subsection{Regime of validity} \label{sec:regime}

Before continuing and analyzing  the evolution of the stress-energy tensor of the dual field theory, let us  pause to comment on the features of the solution, the details of  the expansions and on their  regime of validity.

Let us start from the  metric itself. As stated above, it describes the dynamical process of black hole formation due to  a wave of energy triggered by the forcing function at the boundary, which propagates into the  bulk. For simplicity,  we shall first  refer to the homogeneous limit where  $\vp(v,x) = \phi_{0}(v)$. The metric in this case takes the form  (cf. $ h_{2,0}(v,r,x)$ and $f_{2,0}(v,r,x)$ in \eqref{sol22})
\bal \label{metrichomo}
ds^{2} &=  - \(r^2 -\frac{3\(\dot \phi_0(v)\)^2}{4} - \frac{ \int_{-\infty}^{v} d\tau~\dot\phi_0(\tau)\dddot \phi_0(\tau)}{2 r} \) dv^2+ 2 dv dr \\
& \hspace{5 cm}+ r^2\(1-\frac{\( \dot\phi_0(v)\)^2}{ 8r} \) ( d x^2 + dy^2) \, .
\eal
Away from the energy injection period $0< v< \delta t$, the metric \eqref{metrichomo} has   the AdS-Vaidya form
\be \label{metrichomobis}
ds^2 =  - \(r^2 - \frac{M(v)}{ r} \) dv^2+ 2 dv dr + r^2 ( dx^2  + dy^2)\, .
\ee
The spacetime  is AdS$_4$  for $v\le 0$ ($M(v)=0$); for $v\ge \delta t $, on the other hand, $M(v)$ is a constant $M$ determined by the integral in  \eqref{metrichomo}  and  the geometry approximates a black hole geometry with temperature $T \sim M^{1/3} \sim \frac{\eps^{2/3}}{\delta t}$. (See  \cite{Bhattacharyya:2009uu} for  a more detailed analysis including a discussion of the horizon formation.)
By analogy, in our case, we identify the coefficient of the $1/r$ term in $h(v,r,x)$ as a notion of local  ``temperature'' (by which we do not mean to suggest local equilibrium has yet been established). Once the forcing function has been turned off,  we have
\be \label{mvx}
M(v,x) = -\( C_{2,0}(v,x)  + C_{2,2}(v,x)  +C_{2,4}(v,x) \) \, ,
\ee
with the explicit form of  $C_{2,i}$ given above in \eqref{C20}, \eqref{C22} and \eqref{C24}.

We have obtained the solution in a weak field approximation for the boundary source,  {\it i.e.}  assuming the amplitude of $\vp$ is  small (as indicated by the formal expansion parameter $\eps$). An extensive discussion of  the regime of validity of this approximation  has been presented in  \cite{Bhattacharyya:2009uu}.  The conclusion is that  the amplitude expansion is perturbative in $tT$,\footnote{As we  are mostly interested in the consequences of our analysis on the dual field theory, we focus on the near-boundary region where the field theory time $t$ can be   identified with the bulk time cordinate $v$. The discussion  for the bulk  theory  goes similarly, replacing $tT$ by $vT$. } and it  is therefore reliable as long as  $t \ll 1/T$. In order to capture the evolution for longer times, a resummed perturbation theory is needed, where one expands around AdS-Vaidya rather than AdS. This amounts to working exactly in $T$ and perturbatively in all other appearances of $\epsilon$, which is similar to absorbing temperature-dependent masses in propagators in thermal perturbation theory. At late times, observables approach their thermal values exponentially in $tT$.  In naive (non-resummed) perturbation theory in  $T \sim \eps ^{2/3} / \d t$, the exponential series are truncated to finite order, leading to  polynomial expressions that diverge at late times. Notice, however, that in the homogeneous case  this effect is not yet present up to second order in the $\eps$ expansion, in the sense that there are no  divergent terms at this order. At leading order in $\eps$, the scalar  field  vanishes at late times, while the metric has the black hole form \eqref{metrichomobis} following from \eqref{metrichomo}.

The second expansion we introduced is  the derivative expansion along the inhomogeneous direction. At finite temperature this  corresponds to asking  that the scalar source should be slowly varying in $x$ over distances  set by the local inverse temperature \cite{Bhattacharyya:2009uu}. After the injection of energy has taken place, if $\l$ is the typical length scale  over which the source $\vp(v,x)$ varies, the derivative expansion holds for $ \lambda \gg 1/T$.
However, here the situation is more involved.  For  $v> \d t$, \eqref{mvx} incorporates the general structure of the solution,  which very schematically we can write  up to second order in the derivative expansion as (see \eqref{latetime} for the derivation)
\be
M(v,x) \simeq \frac{\eps^2}{\d t^{3}}\( \l^0  A(x) + \frac{ \d t^2}{\l^2} \left[  B(x) +  \frac{v^2}{ \d t^2}  C(x) + \frac{v}{ \d t}  D(x)   \right] + O(\l^{-4}) \)\, .
\ee
The correction proportional to $B(x)$ can be considered small compared to the leading order term as long as $\l \gg \d t$. Since $1/T  \sim \delta  t / \eps^{2/3}$, the condition $\l \gg \d t$  is automatically implied by  $\l \gg 1/T$ for $\eps \ll 1$. Moreover,  there is a perturbative structure in $v/\l$ that is similar to the effective  perturbative structure in $tT$ of   the amplitude expansion.\footnote{Here this structure appears through the leading time dependent term in the order  $\l^{-2}$ coefficient of $M$. This structure is generically repeated at all orders in the gradient expansion, and it  is due to the presence of the nested time integrals.} In order to have small corrections coming  from the gradient expansion, it  is therefore  also necessary that $v \ll \l $, which  is also automatically satisfied since $1/T \ll \lambda $ and  $v \ll 1/T $.

There is an additional subtlety regarding the period $0 < v< \d t$ during which the energy is injected. This is  apparent in  the analysis of the scalar field solution at order $\eps$ (see \eqref{phi10}-\eqref{phi14})
 \be
  \phi(v,r, x) = \vp(v, x)+ \frac{\dot \vp(v, x)}{r} +  \frac{\vp''(v, x)}{2r^2}  + \dots \, .
 \ee
The first correction involving spatial derivatives  is present only as  long as the scalar forcing function is turned on.  At this order the background is pure AdS,  for which the local temperature is effectively zero and  the condition $\lambda \gg 1/T$ is not satisfied. However, as long as the correction obtained at order $\l^{-2}$ is small compared to the leading $\l^0$ term, one can reliably work within the  derivative expansion.  This  translates into  the condition $1 / r \ll \l$; at smaller values of the radius we lose control over the perturbative description. For large $\lambda$, the part of the $v\approx 0$ region that is uncontrolled is deep into the bulk and can only affect the boundary at late times, so that close to the boundary the perturbative expansion is reliable except at late  times.

All in all, we can conclude that  our expansions are reliable as long as we ensure that  $t \ll 1/T \ll \lambda$.

\setcounter{equation}{0}


\section{Evolution of the boundary stress-energy tensor}
\label{sec:stress}

\subsection{Boundary  stress-energy  tensor} \label{sec:boundaryemt}

The relation between an asymptotically AdS solution for the action \eqref{bmact} and the expectation values of the boundary stress-energy tensor and of the boundary operator associated to the massless bulk scalar has been obtained in Appendix \ref{app:holographicrenormalization}. For  a solution written in Fefferman-Graham coordinates the relation is summarized in equations \eqref{Td3}
\be \label{eq:vevT}
\langle T_{\a\b} \rangle   =  \frac{3}{16 \pi G_{N}} g_{(3),\a\b} \, ,
\ee
and \eqref{O3}
\be \label{eq:vevO}
\langle \O \rangle  = \frac{3}{16 \pi G_{N}} \phi_{(3)}\, ,
\ee
where $\alpha$ and $\beta$ are indices over the boundary coordinates. For notational simplicity we henceforth omit the expectation value symbol.

To read out the stress-energy tensor corresponding to our solution we thus need to perform a change of coordinates to reach the Fefferman-Graham form.
The ansatz we started with has the form
\bal \label{nostra}
ds^{2} &=   -h(v, r,x)dv^{2} +   2dv \( dr + k(v, r,x) dx\)  +  \\[5 pt]
&\hspace{3.5 cm}+f(v,r,x)^2 e^{B(v,r,x)}dx^2+f(v,r,x)^2 e^{-B(v,r,x)}dy^2  \, ,
\eal
and the metric  we obtained up to order $\eps^2$ and $\mu^4$ is schematically of the  form
\bal
h(v, r,x)&= r^2 \( 1 + \frac{ \H_{(2)} }{r^2} +  \frac{ \H_{(3)} }{r^3} +\frac{ \H_{(4)} }{r^4} ) \)   +  O\(\eps^3,\mu^5\) \, ,\\
f(v, r,x)&= r  + \frac{\F_{(1)}}{r} +  \frac{\F_{(2)}}{r^2}+  \frac{\F_{(3)}}{r^3}   +  O\(\eps^3,\mu^5\) \, ,\\
k(v, r,x)&= \K_{(0)}+ \frac{\K_{(1)}}{r} +  O\(\eps^3,\mu^5\) \, , \\
B(v, r,x)&= \frac{\B_{(2)}}{r^2} +  \frac{\B_{(3)}}{r^3} +  \frac{\B_{(4)}}{r^4}  +  O\(\eps^3,\mu^5\) \, ,
\eal
while the scalar field solution can be written as
\be
\phi(v,r,x) = \Psi_{(0)} + \frac{\Psi_{(1)}}{r} + \frac{\Psi_{(2)}}{r^2} + \frac{\Psi_{(3)}}{r^3}+ O\(\eps^3,\mu^5\) \, .
\ee
By comparison with the explicit form of the solution obtained in the previous Section one can identify the expressions of the  various coefficients (see also Appendix \ref{app:change}). The  relevant  ones here are
\be  \label{expmetric}
\begin{aligned}
\dsty \H_{(2)}&=  -\frac{3\(\dot \vp\)^2}{4} -   \frac{\(\vp'  \)^{2}}{4}   \\
\dsty  \H_{(3)}&=    C_{2,0}  +  C_{2,2} + C_{2,4} &\\
\dsty  \F_{(1)}&= -  \frac{\(\dot\varphi \)^2}{ 8}  & \\
\dsty  \F_{(2)}&= - \frac{\dot\varphi \vp''}{12 }  &\\
\end{aligned}\qquad\qquad
\begin{aligned}
 \dsty  \K_{(0)}&=   \frac{1}{2}\dot\varphi \vp' \\
 \dsty  \K_{(1)}&=   \frac{K_{2,1}}{6 } + \frac{K_{2,3}}{6 }  \\
  \dsty  \B_{(2)}&=    \frac{\(\varphi' \)^2}{ 4}  \\
 \dsty  \B_{(3)} &= \beta_{2,2} +\beta_{2,4,3}
\end{aligned}
\ee
and
\be
\Psi_{(0)} =  \varphi \, , \qquad \Psi_{(1)}=  \dot \varphi \, , \qquad  \Psi_{(2)} =  \frac{ \varphi'' }{2} \, , \qquad \Psi_{(3)} = \frac{ \int_{-\infty}^{v}  \vp'''' }{ 8 } \, ,  \label{expscal}
\ee
where we have suppressed the explicit coordinate dependence.

Any asymptotically AdS solution can be brought into Fefferman-Graham form close enough to the boundary ($r \to \infty$).  We therefore look for  a change of coordinates of the form $(v,r,x,y) \to (t,\vr,\z,y)$, such that  the metric  in the new coordinates takes the form
\bal  \label{feffe}
ds^{2} &= \frac{d\vr^{2}}{\vr^2} + \vr^{2}  g_{\a\b}(t, \vr, \z)dx^{\a} dx^{\b} \\
g_{\a\b} (t, \vr, \z)& = g_{(0),\a\b}(t,\z) + \frac{g_{(2),\a\b}(t,\z)}{\vr^{2}} +  \frac{g_{(3),\a\b}(t,\z) }{\vr^{3}}+ \dots \, ,
\eal
where $x^\a= (t,\z,y)$.  Any dependence on $y$,  both in the change of coordinates itself and in the metric, has been excluded a priori for  symmetry reasons.  Notice that the coordinate $\vr$ is simply related to the coordinate used in Appendix \ref{app:holographicrenormalization} by $\vr = 1/ z  $.  We find it  more convenient to work with the coordinate $\vr$ here, so that the boundary $r\to \infty$ corresponds to    $\vr\to\infty$.  In these coordinates,  the scalar field  has the Fefferman-Graham expansion
\be
\phi(t,\vr,\z) = \phi_{(0)}(t,\z) +\frac{ \phi_{(2)}(t,\z)}{\vr^{2} } +  \frac{ \phi_{(3)}(t,\z)}{\vr^{3} } + \dots \, .
\ee
Let us remark that we do not look for an exact change of coordinates, but  we work in a large radius expansion at a sufficiently high order to determine the boundary field theory stress-energy tensor.

For $v\le 0$ the spacetime is pure AdS; the change of coordinates is exact and it reduces to the standard coordinate transformation relating  Eddington-Finkelstein and  Poincar\'e coordinates. In practice we shall look for a change of coordinates such that  $g_{(0),\a\b}  = \eta_{\a\b} = \text{diag}(-1,1,1)$.
Working perturbatively in the radial variable one constructs
the transformation that  brings the scalar field and the  metric in the required form. The details are worked out in Appendix \ref{app:change}, the net result is
\bea \label{cambioo}
v &\to& t -\frac{1}{\vr}  + \frac{v_3}{ \vr^3} + O(\vr^{-4}) \, ,  \nonumber \\
r &\to& \vr \( 1-\frac{\H_{(2)} - 3 v_3}{ 3 \vr^2}  + \frac{ r_3}{\vr^3}  + O( \vr^{-4} ) \)\, ,  \\
x &\to & \z   + \frac{ \K_{(0)} }{ 3\vr^3 } + O( \vr^{-4} ) \, .  \nonumber
\eea
where
\be
v_{3} = \frac{1}{12}\H_{(2)}
\ee
and
 \be
 r_{3} = -\frac{1}{6}\( \H_{(3)} + 4 \F_{(2)}- 4 \dot\F_{(1)}-\frac{1}{3}\dot\H_{(2)} + \frac{2}{3}\K'_{(0)}\)\, .
 \ee

For the scalar field one obtains  in the large $\vr$ expansion
\bea
\phi(t,\vr,\z)  &=&  \vp(t,\z)- \frac{\ddot\vp(t,\z)- \vp''(t,\z) }{2 \vr^2} \\
&&+\frac{ 8 \dddot\vp(t,\z) - 12 \dot \vp''(t,\z) + 3 \int^{t}_{-\infty} d\tau \, \vp''''(\tau,\z)  }{24 \vr^3}  + O\(\eps^{3},\mu^{5},\vr^{-4}\) \, . \nonumber
\eea
It  has exactly the structure expected from the analysis in Appendix \ref{app:holographicrenormalization}: there is no $1/\vr$ term while $\phi_{(2)} = \frac{ \Box_{(0)} \phi_{(0)} }{2}$, as  in \eqref{consd3}.

For what concerns the metric $g_{\a\b}$:   $g_{(0),\a\b}  = \eta_{\a\b} = \text{diag}(-1,1,1)$, while $g_{(2),\a\b}$ can be written as
 \be
 g_{(2),\a\b}=\frac{1}{2}\(\del_{\a} \vp \del_\b\vp -\frac{ \eta_{\a\b} }{4}(\del\vp)^{2}\) \, ,
\ee
 which is the  condition obtained in \eqref{consd3}.
After some algebra (see Appendix \ref{app:change})
\bal
&g_{(3),\a\b}=\frac{1}{3}\times\\[5 pt ]
& \left(
\begin{array}{ccc}
 -2\H_{(3)} &  - \H_{(2)}' - 2 \dot\K_{(0)} + 3 \K_{(1)}  & 0 \\
  - \H_{(2)}' - 2 \dot\K_{(0)} + 3 \K_{(1)}  &- \H_{(3)}  - 3 \B_{(3)} +3 \dot\B_{(2)}   +  \K_{(0)}'
   & 0 \\
 0 & 0  &- \H_{(3)}  + 3 \B_{(3)} -3 \dot\B_{(2)}   -  \K_{(0)}'
\end{array}
\right) \, .
\eal
The last relation in \eqref{consd3} involving $g_{(3)\a\b}$
\be
\del^{\a} g_{(3),\a\b} = \phi_{(3)}\del_{\b}\phi_{(0)}
\ee
is also satisfied, as one can check by explicit computation.

From this analysis the stress-energy tensor and the operator expectation values \eqref{eq:vevT} and \eqref{eq:vevO} can be readily obtained. Renaming the boundary coordinates in a natural way we have
\be \label{stress}
\begin{aligned}
\left.\begin{array}{ll}
T_{tt} =- \frac{2}{16 \pi G_{N}}  \H_{(3)}  & T_{xx}=  -\frac{1}{16 \pi G_{N}} \(\H_{(3)} - 3 \B_{(3)}  + 3 \dot\B_{(2)}  - \K_{(0)}'\) \\
T_{tx}= - \frac{1}{16 \pi G_{N}}     \( \H_{(2)}' +  2 \dot\K_{(0)}  - 3 \K_{(1)} \) & T_{yy} = -\frac{1}{16 \pi G_{N}} \( \H_{(3)} + 3 \B_{(3)}   - 3 \dot\B_{(2)}  +  \K_{(0)}'\)\, .
\end{array}\right.
\end{aligned}
\ee
The explicit expression in terms of the scalar source  can be obtained from \eqref{expmetric}.
For the operator dual to the bulk scalar instead one has
\be
\O = \frac{1}{16 \pi G_{N}}   \frac{ 8 \dddot\vp(t,x) - 12 \dot \vp''(t,x) + 3 \int^{t}_{-\infty} d \tau\, \vp'''' (\tau, x) }{8 }\, .
\ee

In the homogeneous limit ($\vp(t,x) = \phi_{0}(t)$) \eqref{stress} reduces to
\be \label{eq_Thom}
T^{hom}_{tt} = 2 T^{hom}_{xx} =  2 T^{hom}_{yy} = -\frac{1}{16 \pi G_{N}}  \int_{-\infty}^{t} d\tau ~ \dot\phi_0(\tau)\dddot\phi_0(\tau) \, ,
\ee
which is the stress-energy tensor of a perfect conformal fluid in three spacetime dimensions.

\subsection{Evolution following the energy injection} \label{sec:evo}

The explicit expression of the boundary stress-energy tensor in terms of the forcing function $\vp(t,x)$ can be directly  read from \eqref{stress}, although it is quite involved. Some of the main features have been already outlined in the previous Section. We refer to Appendix~\ref{app:boundary} for the complete expressions of the stress-energy tensor up to second order in the scalar source and up to fourth order in the gradient expansion. As an example, we  give here the explicit form of $T_{tt}$,  at second order in the derivative expansion
\bea \label{tttxt}
T_{tt}(t,x) &=&- \frac{1}{16 \pi G_{N}} \left\{  \frac{1}{2} \int_{-\infty}^{t} d\tau~  \Bigg[ 2 \dot\vp(\tau,x)\dddot \vp(\tau,x)  -\( \dot\vp'(\tau,x)\)^2    \right. \nonumber \\[5 pt]
&&   -  4 \dot\vp(\tau,x)\dot\vp''(\tau,x) + 2 \ddot\vp(\tau,x) \vp''(\tau,x) + 2\ddot\vp'(\tau,x)\vp'(\tau,x)  \\[5 pt]
&& \left.+ 2\frac{\del}{\del{x}} \int_{- \infty}^{\tau}ds~\left[ \dot\vp(s,x)\ddot\vp'(s, x) -    \int_{- \infty}^{s} dw~\ddot\vp'(w,x)\ddot\vp(w, x) \right]  \Bigg] \right\} \nonumber \, .
\eea
To make progress in the analysis and to simplify the subsequent numerical treatment, we shall now assume that the temporal dependence of the source function is factorized with respect to the space profile and write
\be \label{eq:factorized}
\vp(t,x) =  u(x) \vp_{0}(t) \, ,
\ee
with
 \be \label{timesource}
\begin{aligned}
\left.\begin{array}{ll}
\vp_0(t)=0 \, , &  t \leq 0 \\
\vp_0(t)=\eps \tilde \varphi_{0}(t)\, , &  0<t< \delta t  \\
\vp_0(t)=0 \, ,  & t \geq \delta t \,.
\end{array}\right.
\end{aligned}
\ee
Substituting this ansatz in \eqref{tttxt} and performing a number of integrations by parts, the corresponding energy density profile takes the form
\bal
 \label{energyshort}
T_{tt}(t,x) &= \frac{1}{16 \pi G_{N}} \left\{   u(x)^2 A(t)  +  \frac{1}{2} \frac{\del^{2}}{\del x^{2}}  u(x)^{2} C(t)  + \frac{1}{2} \frac{\del^{2}}{\del x^{2}} u(x)^{2} D(t)  \right.\\
&\hspace{6 cm }  \left.- \frac{1}{2} \( u'(x)^2 + 4 u(x)u''(x) \) E(t)   \right\}  \, .
\eal
For a source profile that is well localized in time between $0$ and $\delta t$ as in  \eqref{timesource}, the time dependent functions in  \eqref{energyshort} have the following   simple behaviour:
\bal \nonumber
A(t) &= \left\{
\begin{array}{lr}
\alpha(t)=-\int^{ t}_{-\infty} d\tau ~ \dot\vp_0(\tau)\dddot \vp_0(\tau) &\qquad  t < \delta t   \\\\
\bar A = -\int^{\delta t}_{-\infty}  d\tau ~ \dot\vp_0(\tau)\dddot \vp_0(\tau) =\text{const}  &\qquad  t \geq \delta t   \,  \\
\end{array}\right.
\\[3 pt ]
 C(t) &= \left\{
\begin{array}{ll}
\gamma(t)=  \int^{ t}_{-\infty}  d\tau ~ \int^{ \tau}_{-\infty} d w ~ \alpha(w)   &\qquad\qquad  t < \delta t\\ \\
\bar C + (t - \delta t )\bar B + \frac{1}{2} (t^2-\delta t^2)\bar A  &\qquad\qquad  t \geq \delta t  \,    \\
\end{array}\right.\\[3 pt]
D(t) &= \left\{
\begin{array}{lr}
\Delta(t)=-\int^{ t}_{-\infty}  d\tau ~ \ddot\vp_0(\tau) \vp_0(\tau)  &\qquad  t < \delta t\\ \\
\bar D = -\int^{\delta t}_{-\infty} d\tau ~ \ddot\vp_0(\tau)  \vp_0(\tau) =\text{const}   &\qquad  t \geq \delta t  \,    \\
\end{array}\right.
\\[3 pt ]
 E(t) &= \left\{
\begin{array}{ll}
\Upsilon(t) =-\int^{ t}_{-\infty} d\tau ~\( \dot\vp_0(\tau)\)^2  &\qquad\qquad  t < \delta t\\ \\
\bar E = -\int^{\delta t}_{-\infty} d\tau ~ \( \dot\vp_0(\tau)\)^2 = -\bar D &\qquad\qquad  t \geq \delta t  \,    \\
\end{array}\right.
\eal
where
\bea
\bar B  &=&   \int^{\delta t}_{-\infty}\!\!\! d\tau ~ \alpha(\tau) =\text{constant} \, , \\
\bar C &=& \gamma(\delta t) =  \int^{ \delta t}_{-\infty}\!\!\! d\tau ~ \int^{ \tau}_{-\infty}\!\!\! d w ~ \alpha(w) =\text{constant}\, .
\eea
Correspondingly, after all energy has been  injected into the system, $T_{tt}$ has a simple polynomial time dependence
 \bal \label{latetime}
\!T_{tt}( t>\delta t, x)\! & =\!  \frac{1}{16 \pi G_{N}}    \left\{    u(x)^2 \bar A   +  \frac{1}{2}   \frac{ \del^{2} }{ \del x^{2} }  u(x)^{2} \!\!\( \bar C + (t - \delta t ) \bar B + \frac{1}{2} (t^2- \delta t^2) \bar A  \) \right.   \\
&\hspace{3.5 cm } \left. + \frac 1 2 \left( 3 (u'(x))^2 + 6 u(x) u''(x) \right) \bar D  \right\}  \, .
\eal
As already pointed out in Section \ref{sec:regime}, this structure comes from the nested integrals over time and gives an effective expansion in $t$ times the inverse length scale of spatial variations. The structure, which at first seems to lead to a divergent profile at arbitrary late time, has to do with our approximations, which fail for large times.
Notice there is no contradiction with energy conservation.  In fact time dependent terms in  \eqref{latetime} involve spatial derivatives and their contribution to the total energy is vanishing. (The same is true at fourth order in the gradient expansion,  as  one can  check from the explicit expression of $T_{tt}$ in \eqref{Ttt4order}. After the injection of energy has completed ($t \geq \delta t$),  time dependent contributions to $T_{tt}$ correspond to  terms involving nested time integrals in \eqref{Ttt4order}, all of which are total spatial derivatives.)

In order to study the stress-energy tensor in more detail and to better understand its evolution, we shall use a Gaussian profile to  mimic the time dependence of the source
\be \label{fattorizza}
\varphi(t, x) =   \eps  u(x) e^{-\frac{\left(t -\nu \right)^2}{\sigma^2}}\, .
\ee
Such a replacement retains the main featurlarger thanes of the evolution discussed above:  when  $t -\nu$ is  large enough,   the scalar source  can be considered to vanish for all practical purposes.

Consider first the simple case where the spatial profile $u(x)$ has the form of a Gaussian on top of a homogeneous background
\be \label{sorgentehomo}
\varphi(t, x) = \eps  \(1+ e^{- \mu^2 x^2}\)  e^{-\frac{\left(t -\nu\right)^2}{\sigma^2}}\, .
\ee
Here we insert  the explicit factor $\mu \sim 1/ \lambda$ in the space dependence of the source  in order to implement  more easily the slowly varying regime. We have considered a homogeneous component in the source field, which gives a non-zero energy density everywhere. In this way the local ``temperature'' is everywhere non-vanishing and we can  keep under good computational control both the weak field and the gradient expansion.

\begin{figure}[h]
\begin{center}
\includegraphics[width = 0.42 \textwidth]{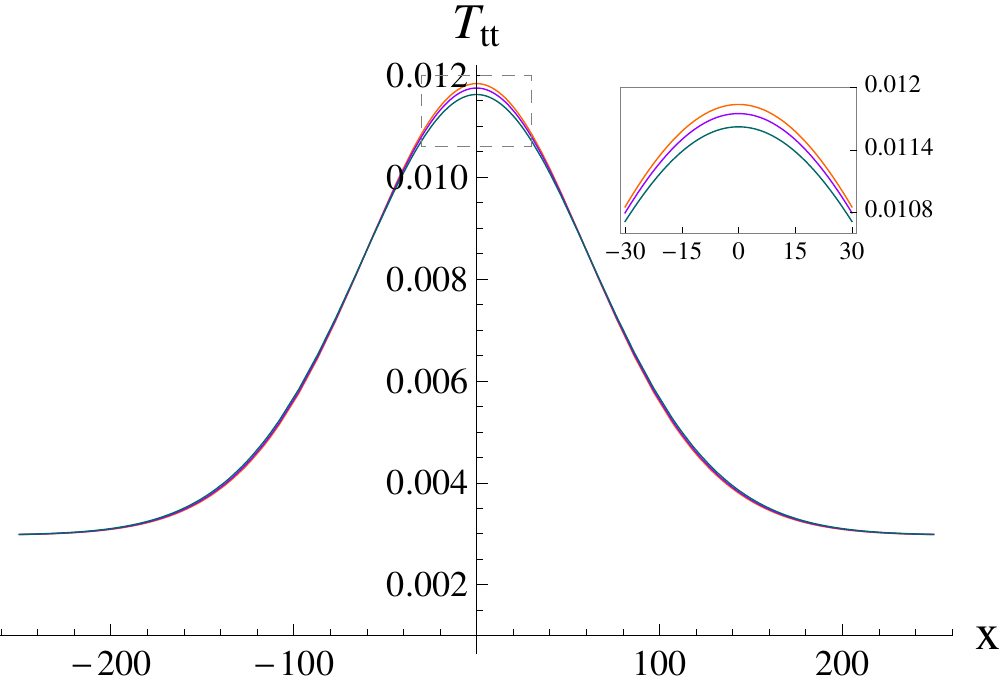} \hfill \includegraphics[width = 0.42 \textwidth]{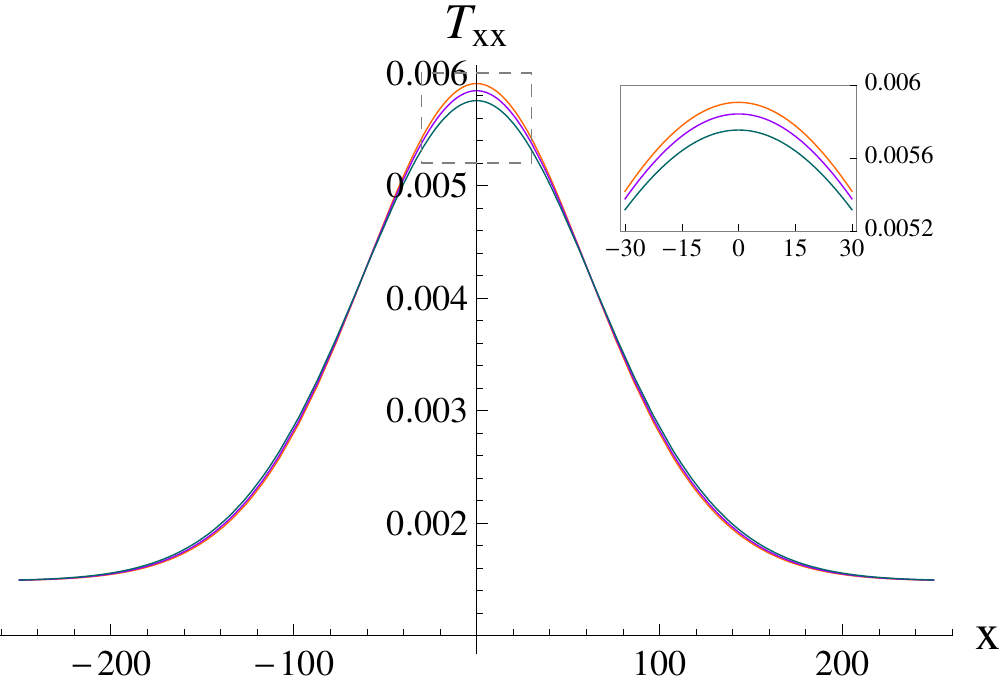}
\includegraphics[width = 0.42 \textwidth]{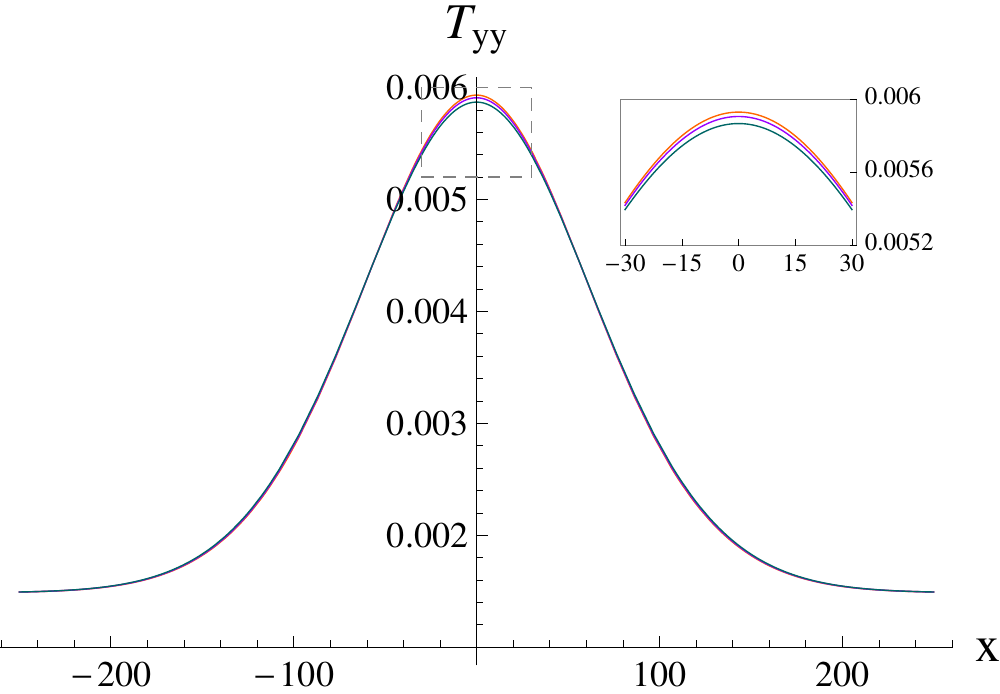}  \hfill   \includegraphics[width = 0.42 \textwidth]{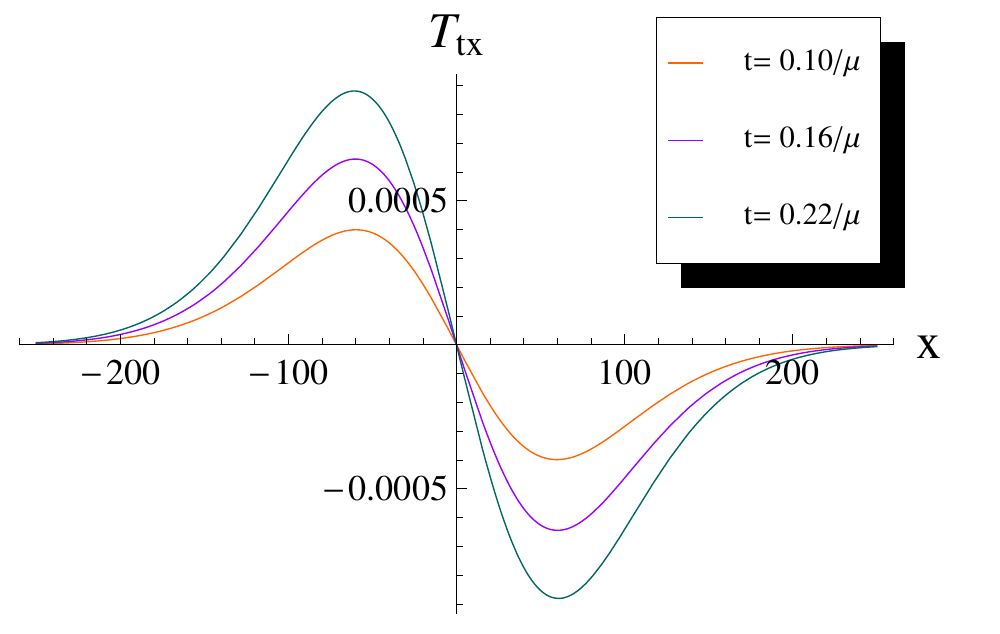}
\caption{Spatial dependence of  $T_{tt}$, $T_{xx}$, $T_{yy}$, $T_{tx}$ for a Gaussian temporal and spatial profile of the source \eqref{sorgentehomo},  for $\nu =0.5$, $\sigma^2 = 0.1$,  $\mu = 0.01$ and $\eps  = 0.005$. The plots show three different instants in the evolution of the profiles (rescaled by $16 \pi G_N$) at   fourth order in the gradient expansion. As  time increases,  $T_{tt}$, $T_{xx}$ and $T_{yy}$ decrease in the region around  $x=0$ and increase  for large $|x|$; while the maximal amplitude of $T_{tx}$ increases in time. A general tendency towards  spatial homogeneity is shown both for $T_{xx}$ and $T_{yy}$, which is however faster for the former. }\label{fig:Ttt}
\end{center}
\end{figure}

In Figure~\ref{fig:Ttt} (top-left panel) we report a sample plot of the $T_{tt}$ component of the stress-energy tensor  for $\nu =0.5$, $\sigma^2 = 0.1$,  $\mu = 0.01$ and $\eps  = 0.005$.  In order to stop  the evolution at a time which is reliably within the range of validity of  our approximations, we examine the effects of sub-leading  contributions  in the gradient expansion.  For a fixed spatial interval, the expansion is valid till times at which the sub-leading contributions become comparable to the leading order result. In line with our general discussion of the regime of validity of the amplitude expansion, we also impose that $t\ll 1/T$, with $T$  the local  ``temperature'' defined in the previous Section by analogy to the equilibrium case.

From the point of view of  $T_{tt}$, the evolution seems to point towards  homogenization, although we cannot follow the evolution for arbitrarily large times. The same   qualitative conclusion seems to hold looking at  the evolution of the other diagonal components. In the top-right and bottom-left panels of Fig.~\ref{fig:Ttt}, the $T_{xx}$ and $T_{yy}$ components of the stress-energy tensor are plotted for the same values of the  parameters, while the only non-vanishing off-diagonal component $T_{tx}$ is plotted in the bottom-right panel.
As opposed to the diagonal components, $T_{tx}$  grows in time, departing from a homogeneous configuration. However, the off-diagonal component of the boundary stress-energy tensor also gets  a contribution from the local  boost velocity,  and the corresponding profile seems to be qualitatively compatible with this interpretation: Intuitively, a lump of energy with a density profile as in Fig.~\ref{fig:Ttt} would start spreading outward and the local velocity with respect to the background is maximal where the gradient of the energy density is large.

\begin{figure}[h]
\begin{center}
\includegraphics[width = 0.42 \textwidth]{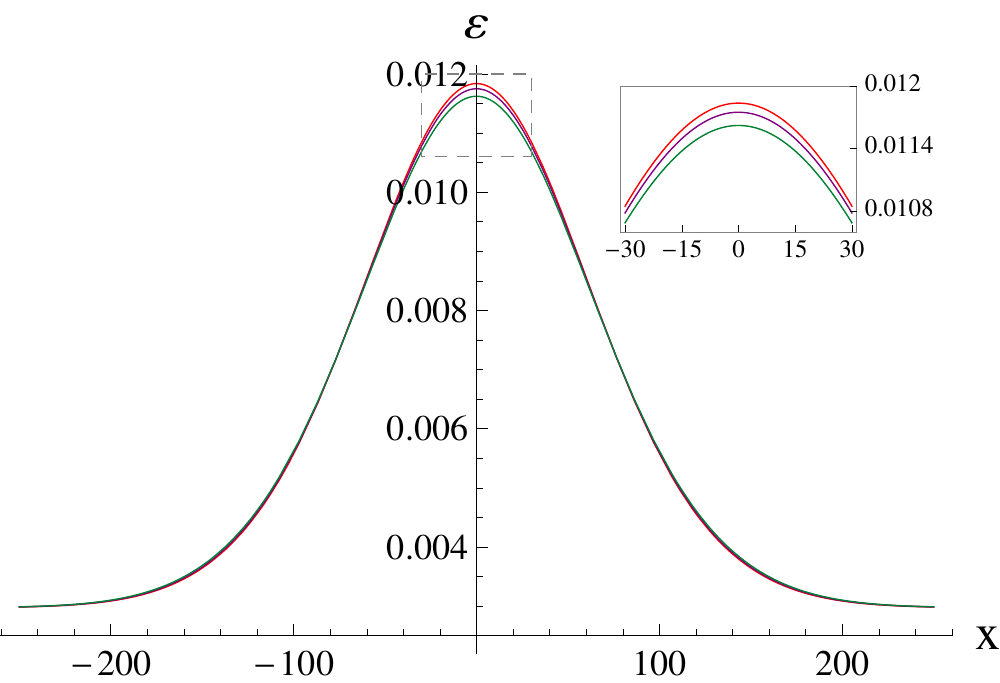}
\hfill
\includegraphics[width = 0.42 \textwidth]{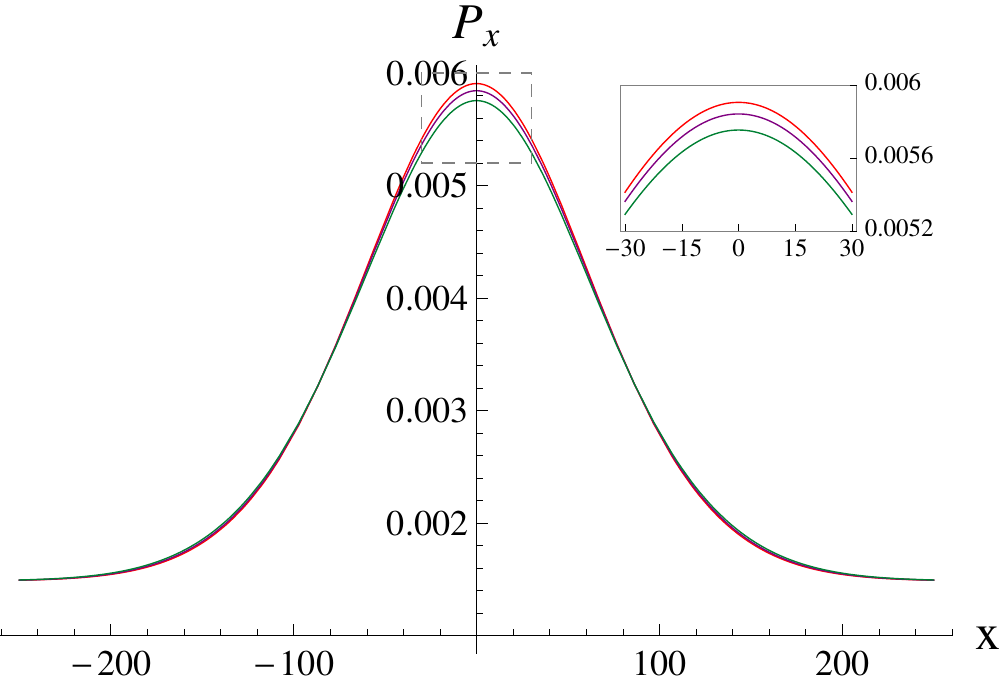}  \\
\includegraphics[width = 0.42 \textwidth]{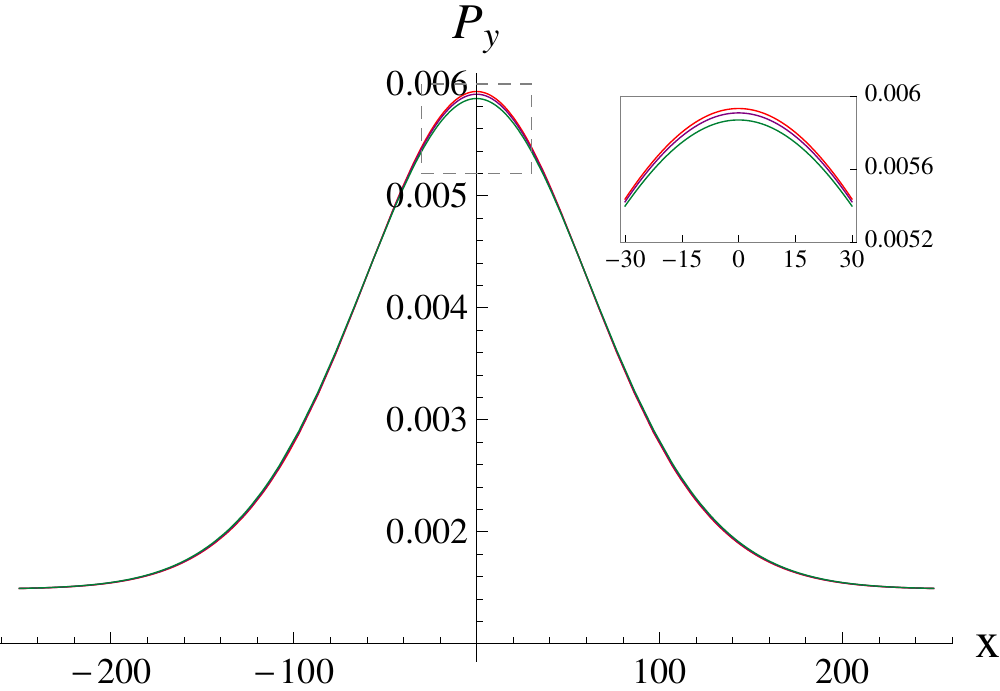}
\hfill
\includegraphics[width = 0.42 \textwidth]{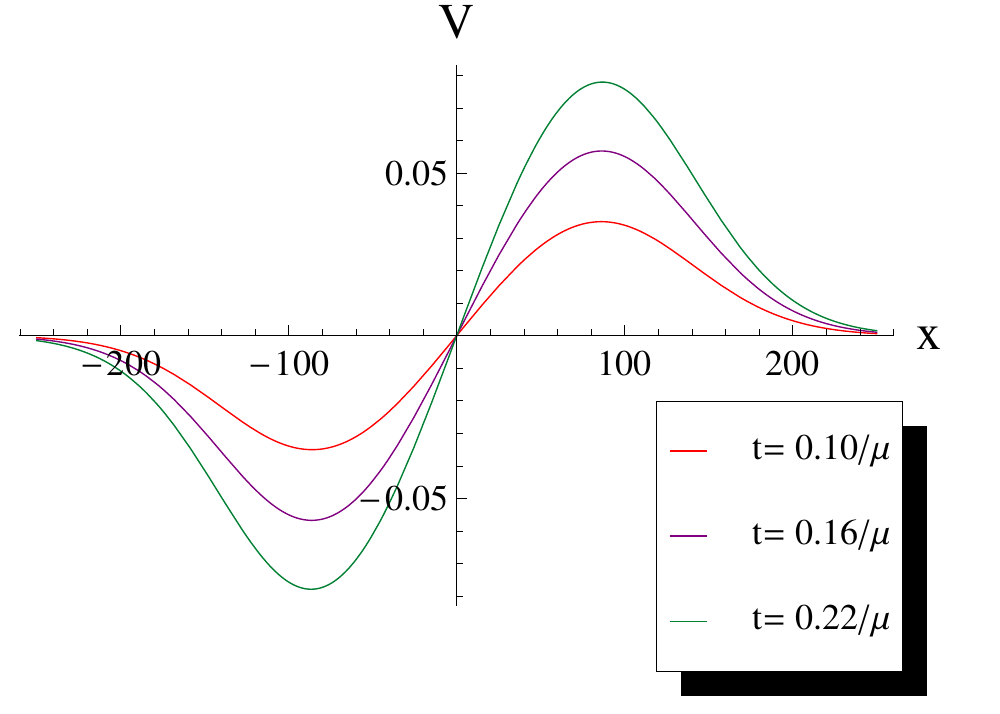}
\caption{The energy density in the rest frame $\varepsilon =p_x +p_y$, pressures $p_x$ and  $p_y$ computed in \eqref{eq:epsilon} (and rescaled with $16 \pi G_N$)  for the solution with  $\nu =0.5$, $\sigma^2 = 0.1$,  $\mu = 0.01$ and $\eps  = 0.005$. As  time increases,  $\varepsilon$, $p_x$ and $p_y$ decrease in the region around  $x=0$ and increase  for large $|x|$.  On the bottom-right panel the plasma  local velocity $V$, which maximal amplitude increases in time. }\label{fig:palpha}
\end{center}
\end{figure}

\begin{figure}[h]
\begin{center}
\includegraphics[width = 0.5 \textwidth]{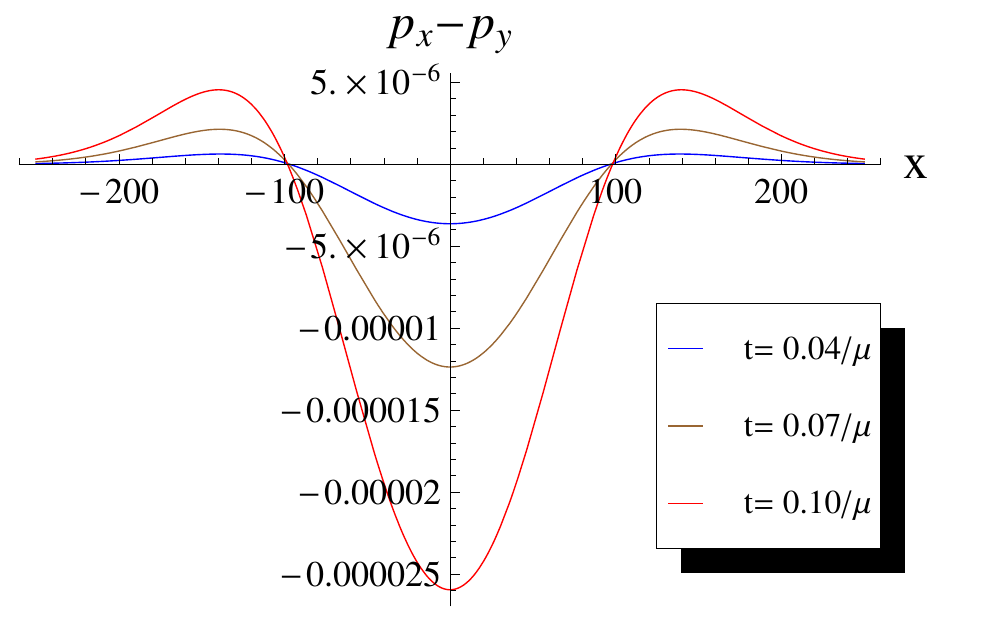}
\caption{The  spatial profile of the pressures anisotropies $p_x-p_y$ (rescaled by $16 \pi G_N$) for the pressures  plotted in Fig.~\ref{fig:palpha}. The plot is limited to shorter times as compared to those considered in  Fig.~\ref{fig:palpha} to ensure that the corrections to $p_x$ and $p_y$ coming from higher orders in the derivative expansion are negligible compared to the pressure anisotropies at leading order. In the plot time increases from the top curve down.}\label{fig:difference}
\end{center}
\end{figure}

To have a  better understanding of the situation it is  convenient to perform  the boost  that locally brings the stress-energy tensor in the diagonal form  $T^{\a\b} = \textrm{diag}(\varepsilon, p_x,p_y)$.
This  boost in the $x$ direction with rapidity $\alpha$  relates the components in the two frames as
\bal \label{eq:epsilon}
&T^{tt}= \varepsilon \cosh^2 \alpha + p_x \sinh^2 \alpha \qquad\quad
T^{tx}=- \frac{\varepsilon +p_x}{2} \sinh 2 \alpha& \\
&T^{xx}= \varepsilon \sinh^2 \alpha + p_x \cosh^2 \alpha \qquad\quad
T^{yy}=p_y\, . &
\eal
The local velocity of the plasma in the stationary frame is $V = - \tanh \alpha$, where
\be \label{eq:alp}
\tanh 2 \alpha = -  \frac{2T^{tx}}{T^{tt} + T^{xx}}\, .
\ee
Using \eqref{eq:epsilon} and \eqref{eq:alp}, we can plot in Fig.~\ref{fig:palpha} the energy density and pressures of the fluid in the local rest frame, as well as  its local velocity $V$.
The energy density and pressures tend to flatten out as expected, while the pressures anisotropies still build up in this phase as shown in Fig.~\ref{fig:difference}.
In order to ensure that the gradient expansion is reliable, in Fig.~\ref{fig:difference} we restrict the time range such that third (fourth) order terms in the gradient expansion are at least two orders of magnitude smaller than first (second) order terms. In Figures~\ref{fig:Ttt} and~\ref{fig:palpha} we relax the time range such that the corrections coming from the subleading order in the gradient expansion should be one order of magnitude smaller than the leading order result, because otherwise the time evolution of the various quantities would not be visible in the plots.
Similar evolution was found for instance by Chesler and Yaffe in \cite{Chesler:2008hg}, where it was observed that the generation of anisotropies  in the stress-energy tensor due to an anisotropic source continues even after  the end of energy  injection. The regime where the anisotropy is built up should then be followed at later times by an evolution towards isotropy. As expected from the discussion in Section~\ref{sec:regime}, this relaxation dynamics is not visible in naive perturbation theory and should start over timescales set by the inverse temperature. Also,  the local boost velocity is qualitatively consistent with that of  a lump of fluid that spreads out. It decreases for large values of $|x|$, consistently with  expectations for a lump of fluid with higher density than the surrounding homogeneous medium.

\setcounter{equation}{0}

\section{Comparison with free streaming and hydrodynamics}\label{sec:comparison}

In order to gain some insight into these results, we next compare them with the evolution of the stress-energy tensor in a free-streaming model as well as in hydrodynamics.

\subsection{Anisotropic free streaming}\label{sect:freestreaming}

To compare the evolution we find using AdS/CFT with a simple model of free-streaming particles, we construct a kinetic model of the injected energy distribution.
Our model is obtained by assuming that the distribution is massless noninteracting dust, composed of particles moving at the speed of light. In terms of the phase space distribution $f(t, \vec x, \vec k)$, the  components of the stress-energy tensor are given by
\be
T^{\alpha\beta}(t, \vec x) = \int d^2k \frac{k^\alpha k^\beta}{k^t} f(t, \vec x, \vec k) \,.
\label{eq:Tmunu-streaming}
\ee
If we assume the phase space distribution at some time $t=\delta t$ to have the factorized form $f(\vec x, \vec k) = n(\vec x) F(\vec k) = n(x) F(k)$, then at a later time
\be \label{eq:f}
f(t, \vec x, \vec k) = n(\vec x-\vec v (t-\delta t)) F(\vec k) = n(x-v_x (t-\delta t)) F(k),
\ee
where $\vec v=\vec k/k^t$ is the particle velocity, and $F(k)$ only appears in \eq{eq:Tmunu-streaming} in an overall normalization factor.
It turns out that the specific form of the function $F(k)$ is irrelevant, what matters is that all particles move at the speed of light.
When comparing with our AdS/CFT results, we start the free streaming evolution just after the energy injection has ended, namely at $t =\delta t$, and choose $n(x)$ and $F(k)$ such that the energy density in the laboratory frame coincides with that computed using AdS/CFT.

We can introduce polar coordinates and write
\be
k_x = k\cos\phi, \quad k_y = k\sin\phi, \quad v_x = \cos\phi, \quad v_y = \sin\phi .
\ee
One easily finds that $T^{xy}=0$, which means that for this particular choice there is no effective flow developing. This does not have to be the case when the $k$ and $x$ dependence are not factorized. The non-vanishing components of $T^{\alpha\beta}$ are:
\ba \label{eq:Tfreestreaming}
T^{tt}(x,t) &=& \frac{\varepsilon_0}{2\pi} \int_0^{2\pi} d\phi\,\, n(x-(t-\delta t)\cos\phi) ,
\nn \\
T^{tx}(x,t) &=& \frac{\varepsilon_0}{2\pi} \int_0^{2\pi} d\phi\, \cos\phi\,\, n(x-(t-\delta t)\cos\phi) ,
\nn \\
T^{xx}(x,t) &=& \frac{\varepsilon_0}{2\pi} \int_0^{2\pi} d\phi\, \cos^2\phi\,\, n(x-(t-\delta t)\cos\phi) ,
\nn \\
T^{yy}(x,t) &=& \frac{\varepsilon_0}{2\pi} \int_0^{2\pi} d\phi\, \sin^2\phi\,\, n(x-(t-\delta t)\cos\phi) ,
\ea
where the common factor is given by the initial average energy per particle:
\be
\varepsilon_0 = 2\pi \int_0^{\infty} k^2 dk F(k) .
\ee
The tracelessness of $T^{\alpha\beta}$ is easily seen from these expressions.

Notice that at $t =\delta t$, the stress-energy tensor in \eqref{eq:Tfreestreaming} is isotropic, while in \eqref{stress} anisotropies start to develop already during the energy injection. However, their relative amplitude only starts to build up considerably after a time $\delta t$, and therefore the difference between the two stress-energy tensors  at $t=\delta t $ is not significant.

\begin{figure}[h]
\begin{center}
\includegraphics[width = 0.42 \textwidth]{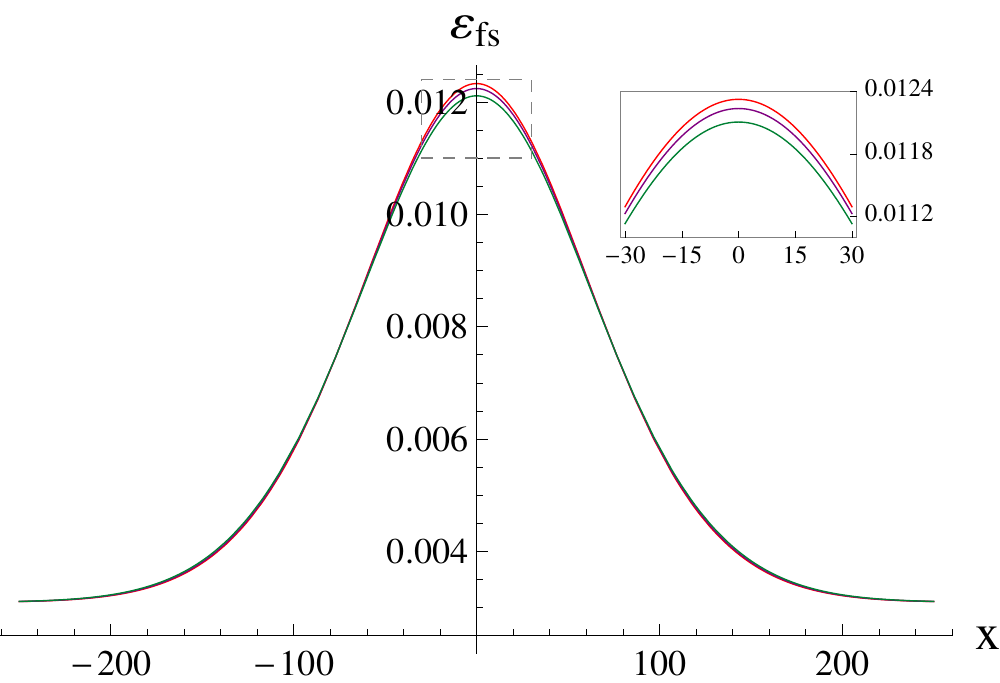}
\hfill
\includegraphics[width = 0.42 \textwidth]{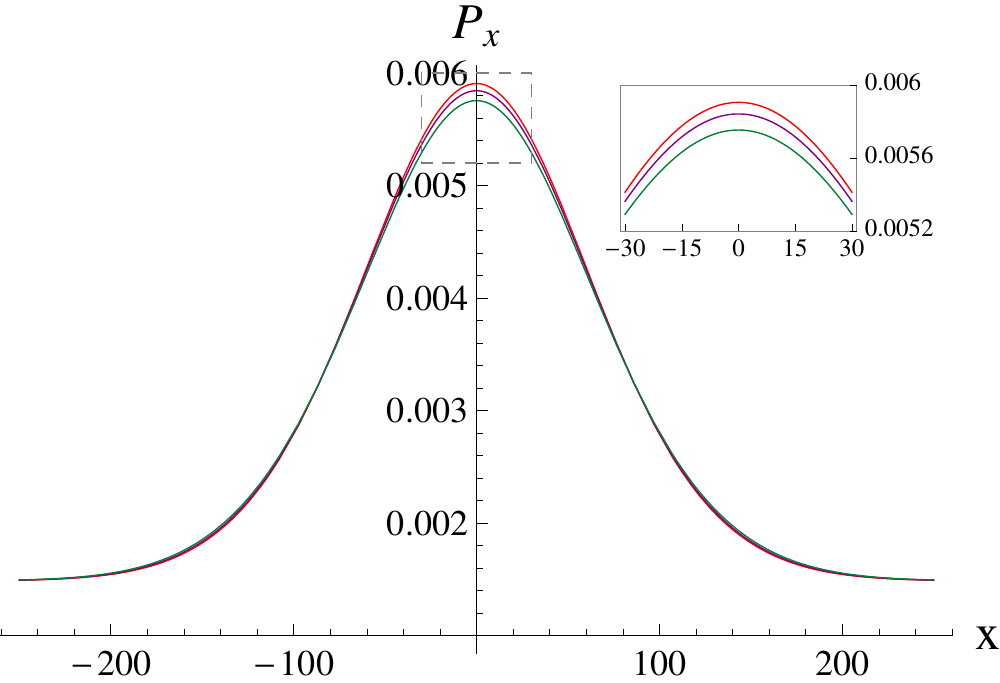}  \\
\includegraphics[width = 0.42 \textwidth]{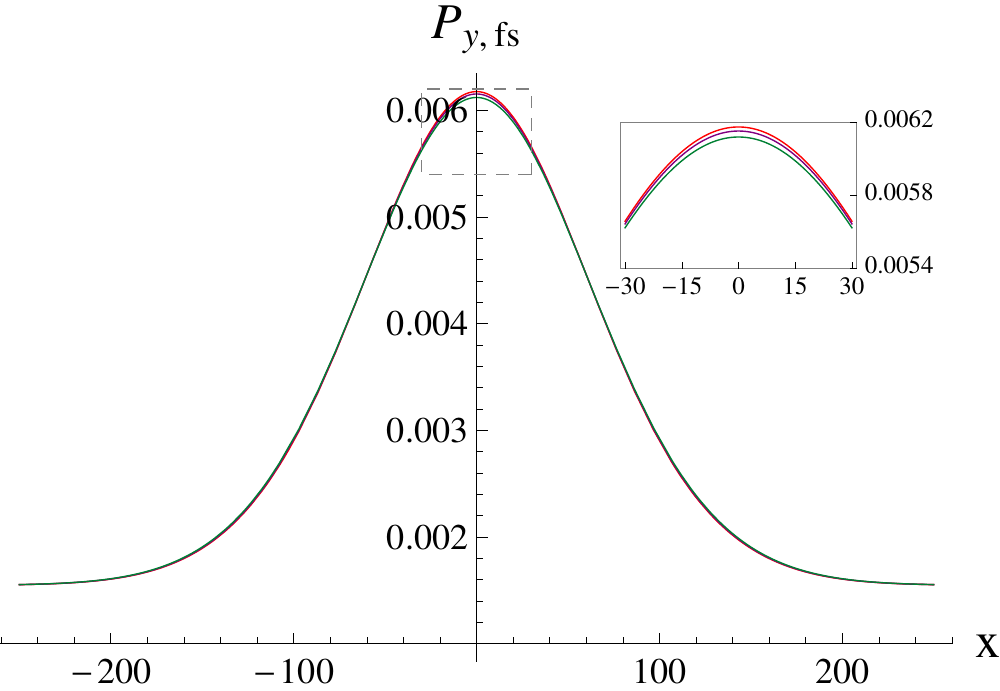}
\hfill
\includegraphics[width = 0.42 \textwidth]{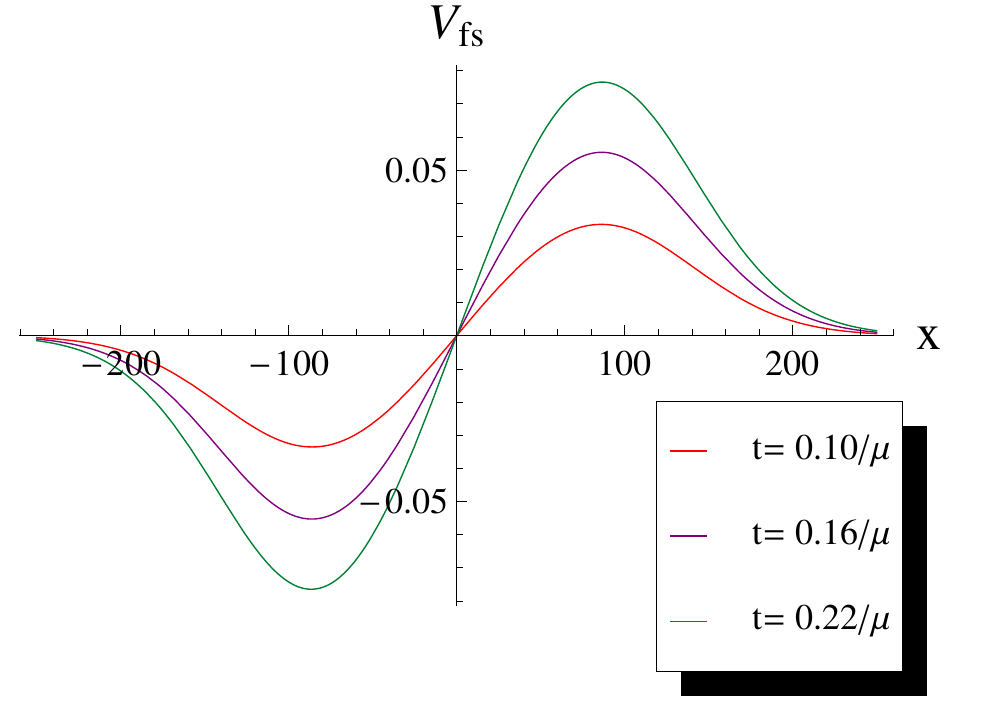}
\caption{The energy density in the rest frame $\varepsilon_{\rm fs} =p_{x, {\rm fs}} +p_{y, {\rm fs}}$, pressures $p_{x, {\rm fs}} $ and  $p_{y, {\rm fs}}$ (rescaled with $16 \pi G_N$) and local boost velocity $V_{\rm fs}$ within free streaming. The match of the free streaming and the AdS/CFT energy densities in the laboratory frame has been performed at $t = 0.009/\mu$, when about  $96 \%$ of the energy has been injected.
The evolution closely resembles that obtained with the AdS/CFT correspondence and plotted in Fig.~\ref{fig:palpha}. }\label{fig:palpha_fs}
\end{center}
\end{figure}

\begin{figure}[h]
\begin{center}
\includegraphics[width = 0.5 \textwidth]{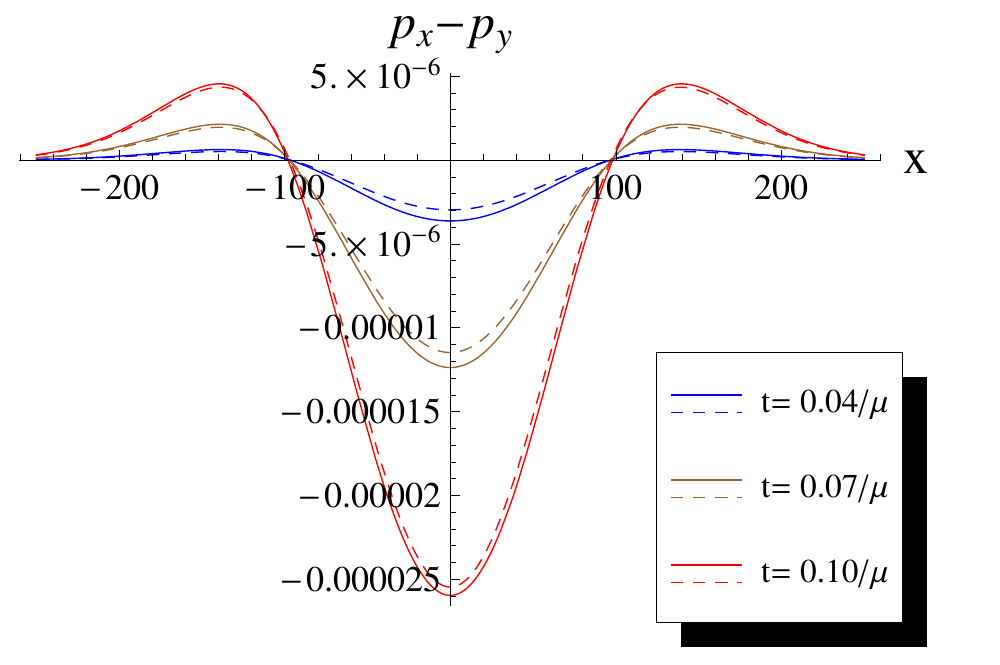}
\caption{The pressure anisotropies (rescaled with $16 \pi G_N$) obtained within anisotropic free streaming evolution and following from Fig.~\ref{fig:palpha_fs} (in dashed) closely approximate those obtained with the AdS/CFT computation (solid lines) in the range of times we are able to explore.}\label{fig:anisotropies}
\end{center}
\end{figure}

In Fig.~\ref{fig:palpha_fs}, we plot the  rest frame energy density $\varepsilon$, the pressures $p_x$ and  $p_y$ and the local velocity $V$ of the plasma during a free streaming evolution.
The curves closely resemble those obtained from the holographic computation in Fig.~\ref{fig:palpha}, although the maximal amplitude of the energy density and pressures decreases more slowly under anisotropic free streaming. Nevertheless, the pressure anisotropy in the local rest frame matches well that obtained within AdS/CFT, as shown in Fig.~\ref{fig:anisotropies}.
%

\subsection{Viscous hydrodynamics}\label{sect:viscous}

Hydrodynamics is a low-energy description that is not meant to be valid on time scales smaller than the inverse local temperature. Nevertheless, viscous hydrodynamics turns out to give surprisingly good results in several homogeneous holographic thermalization setups \cite{Chesler:2009cy,Beuf:2009cx,Heller:2011ju}. Motivated by this, we now compare our AdS/CFT results with what one would obtain by naively applying the formulas of viscous hydrodynamics. In fact, in \cite{Bhattacharyya:2009uu} it has been suggested that the anisotropies present after the injection of energy  could be interpreted in terms of first order viscous hydrodynamics, for which the stress-energy tensor is completely expressed in terms of the local temperature and of the local velocity of the fluid (a  review of the basic ingredients can be found for instance in \cite{Hubeny:2011hd}).

Given the local velocity $V(t,x)$ and the energy density in the local rest frame $\varepsilon(t,x)$, we can ask what the stress-tensor would be if hydrodynamics were valid at a given time $t$. Agreement with the stress tensor we computed using AdS/CFT would be a necessary condition for the validity of a hydrodynamical description from time $t$ onwards.

\subsubsection{First order viscous hydrodynamics}

First order hydrodynamics requires the equation of state and the shear and bulk viscosities, which can be found, for instance, in \cite{Hubeny:2011hd,Bhattacharyya:2008mz} for the three-dimensional CFT under consideration. The first-order hydrodynamical stress tensor reads
\be \label{eq:Tvisc}
T_{\rm viscous}^{\a\beta} =  \( \varepsilon + p_{\rm ideal } \) u^\a u^\beta + p_{\rm ideal } \, \eta^{\a\beta} + \Pi^{\a\beta}_{(1)}\, .
\ee
Here $p_{\rm ideal }=\varepsilon/2$ and $u^\alpha$ is the local three-velocity of the fluid (which we determine from $V(t,x)$). The first order viscous contributions to the stress-energy tensor in flat three-dimensional space are given by \cite{Hubeny:2011hd}
\be
\Pi^{\a\beta}_{(1)} \equiv - 2 \eta \sigma^{\a \beta} - \zeta \theta P^{\a\beta}\,,
\ee
where $P^{\a \beta} \equiv u^\a u^\beta + \eta^{\a\beta}$ is the projector onto space in the local fluid rest frame,
\be
\sigma^{\a\beta} \equiv P^{\a\rho} P^{\beta\sigma} \left(\del_{(\rho}u_{\sigma)} - \frac 1 2 P_{\rho\sigma} \theta \right)
\ee
is the fluid shear tensor and $\theta \equiv \del_\rho u^\rho$ is the expansion. The shear viscosity $\eta$ and bulk viscosity $\zeta$ are \cite{Hubeny:2011hd,Bhattacharyya:2008mz}
\be \label{eq:minimalshear}
\eta = \frac{1}{16 \pi G_N} \left(\frac{4}{3} \pi T\right)^2\,, \qquad \zeta = 0\, ,
\ee
where we define a local ``temperature'' $T(t,x)$ by using the thermodynamic relation that would be valid in equilibrium in the local rest frame  \cite{Bhattacharyya:2008mz},
\be
\varepsilon=  \frac{2}{16\pi G_N} \(\frac{4}{3} \pi T\)^3\, .
\ee

To compare with the stress tensor we have worked out, we use the fluid velocity $V = -\tanh \alpha$ determined in the previous Section for a single Gaussian scalar profile.

\begin{figure}[h]
\begin{center}
\includegraphics[width = 0.5 \textwidth]{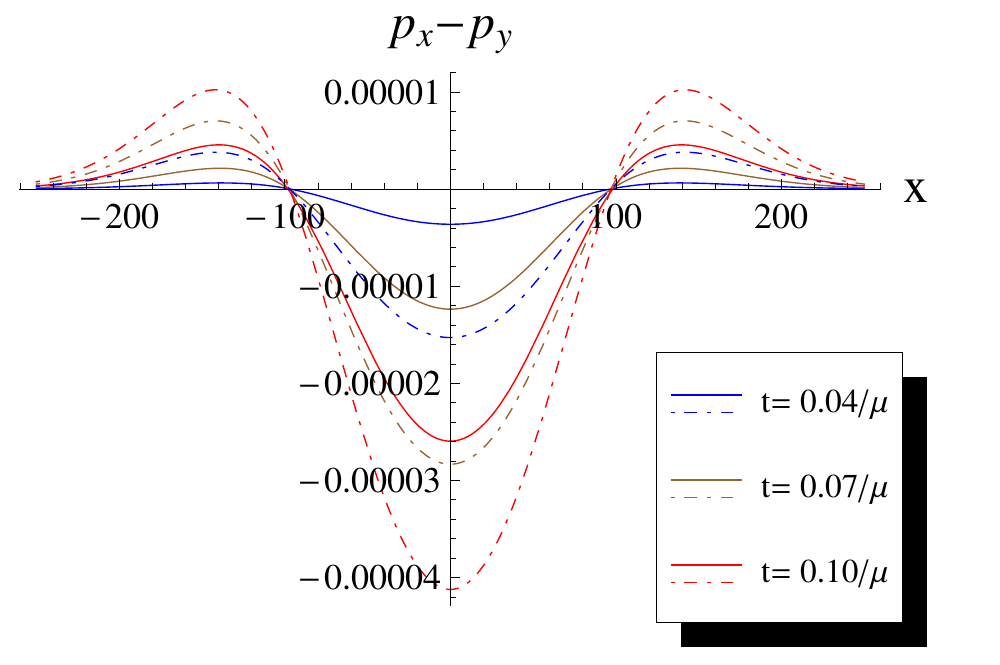}
\caption{Comparison of the  spatial profile of  the pressure anisotropies $p_x - p_y$ (rescaled with $16 \pi G_N$) of Fig.~\ref{fig:difference}  with that resulting from first order hydrodynamics. The pressures anisotropies computed in the gradient expansion (solid curves) are everywhere smaller in amplitude than those computed in first order viscous hydrodynamics (in dot-dashed).}\label{fig:difference_homo}
\end{center}
\end{figure}

Determining the pressure components $p_{x,{\rm hydro}}$ and $p_{y,{\rm hydro}}$ in the local rest frame, the first order pressure anisotropy reads
\be
p_{x, {\rm hydro}} - p_{y, {\rm hydro}} = - 2 \eta \theta.
\ee
In Fig.~\ref{fig:difference_homo}, this is compared with the pressure anisotropies computed in the gradient expansion. The two have amplitudes of the same order of magnitude and similar shape, although they do not coincide. In particular the ones computed in the gradient expansion are everywhere smaller in amplitude than the hydrodynamics ones.
%

\subsubsection{Second order hydrodynamic corrections}\label{sect:viscous2}

Since first order viscous hydrodynamics did not give a sufficiently accurate description of the time-evolution of the initial inhomogeneities, we move to second order formalism.

Second order hydrodynamics for relativistic conformal fluids was derived in \cite{Baier:2007ix, Bhattacharyya:2008jc}, providing a nonlinear generalization of M\"uller-Israel-Stewart theory, used in modeling heavy ion collisions.
The relevant second order contribution to the dissipative part of the stress tensor is (see eqn (3.11) of \cite{Baier:2007ix})
\be\label{second}
\Pi^{\alpha \beta}_{(2)} =  \eta \tau_{\Pi} \left[  \mbox{}^{\langle} D\sigma^{\alpha \beta \rangle} + \half \sigma^{\alpha\beta} \theta \right] + \cdots
\ee
where $``+\cdots"$ refers to terms that are not relevant here, as they either contain the Riemann/Ricci tensor or terms that vanish for irrotational flow. In  the above,
$D$ is the directional derivative along the 3-velocity:
\be
D \equiv u^\alpha \partial_\alpha = -u_t\partial_t+u_x\partial_x \  ,
\ee
while $\langle \rangle$ denotes the transverse traceless part. The new parameter $\tau_{\Pi}$ is a relaxation timescale, which for a strongly coupled conformal fluid  depends on harmonic numbers \cite{Bhattacharyya:2008mz} and in 2+1 dimensions
\be\label{relaxtime}
\tau_{\Pi} = \frac{3}{4\pi T}\left[ 1 + \frac{1}{3} {\rm Harmonic}\left(-\frac{1}{3}\right)\right]  \approx \frac{0.180}{T}.
\ee
 The central feature of the second order theory
is that now hydrodynamics becomes causal, with the discontinuities/inhomogeneities propagating at finite velocity \cite{Baier:2007ix, Bhattacharyya:2008jc}
\be\label{vdisc}
v_{disc} = \sqrt{\frac{\eta}{ \tau_{\Pi} (\epsilon+ p_{\rm ideal})}} = \sqrt{\frac{1}{3 + {\rm Harmonic}\left(-\frac{1}{3}\right)}} \approx 0.665 \ .
\ee

For the second order term in  (\ref{second}) we need to evaluate first
\be\label{Dsigma}
D \sigma^{\alpha \beta} = 2(D\theta )M^{\alpha \beta} + 2\theta DM^{\alpha \beta} ,
\ee
where
\be
M^{\alpha \beta} \equiv \frac{ \partial^{\langle \alpha}u^{\beta\rangle} }{\theta},
\ee
and then construct its transverse traceless projection $ \mbox{}^{\langle} D\sigma^{\alpha \beta\rangle}$. The first term in (\ref{Dsigma}) is already transverse and traceless. By an explicit calculation one can verify that the second term projects to a null matrix. The second order contribution to the  dissipative part of the stress tensor thus becomes
\be
 \Pi_{(2)}^{\alpha \beta}  = 2 \eta \tau_{\Pi} \(D\theta +\half \theta^2 \)M^{\alpha \beta} \ ,
\ee
and in total then
\be
\Pi^{\alpha \beta} \equiv  \Pi_{(1)}^{\alpha \beta} + \Pi_{(2)}^{\alpha \beta} = 2\eta \(-\theta + \tau_{\Pi} \(D\theta + \half \theta^2 \)\)M^{\alpha \beta} \ .
\ee
Explicitly in our case the pressure anisotropies in the local rest frame read:
\be
p_{x, {\rm hydro}} - p_{y, {\rm hydro}} = 2 \eta \(- \theta +  \tau_{\Pi} \( D\theta + \frac 1 2 \theta^2\) \)\,.
\ee
The first and second order viscous contributions to the pressure anisotropies have opposite signs and the latter dominate over the first at early times. The dominant term at early  times is in fact  $2 \eta \tau_\Pi u^t \del_t \theta$.

\begin{figure}[h]
\begin{center}
\includegraphics[width = 0.5 \textwidth]{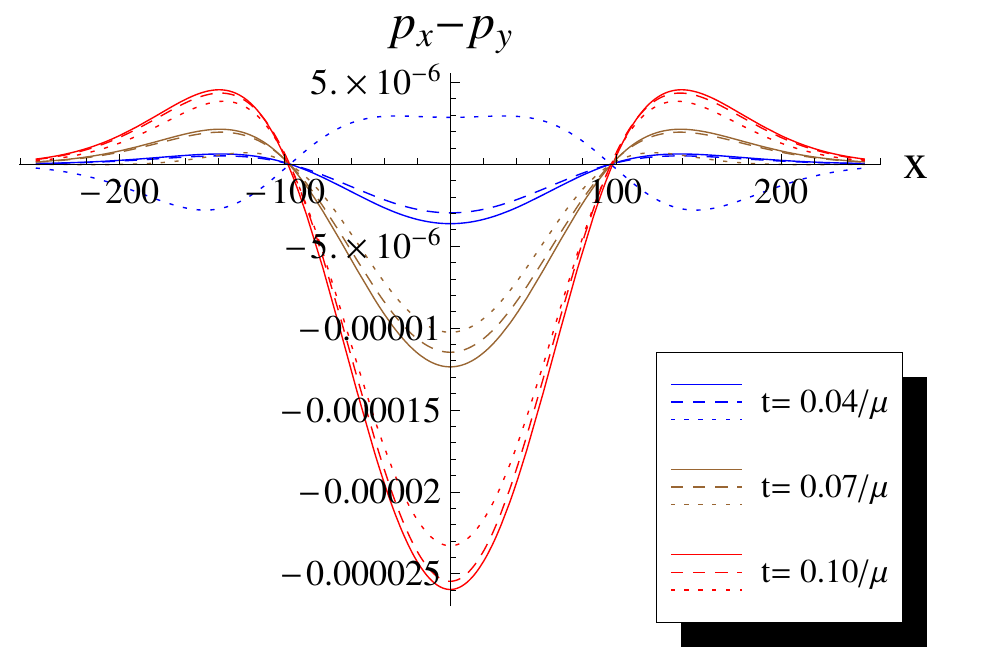}
\caption{The  spatial profile of  the pressures anisotropies $p_x - p_y$ (rescaled by $16 \pi G_N$) of Fig.~\ref{fig:difference} (solid lines) compared with those resulting from second order hydrodynamics (dotted) and free streaming (dashed). The addition of the second order correction drives the pressure anisotropy in second order hydrodynamics to a good quantitative matching with the gradient expansion near the upper time $t=0.1/\mu$.}\label{fig:anisotropies2}
\end{center}
\end{figure}

As we saw in Figure~\ref{fig:difference_homo} in the previous section, in first order hydrodynamics the pressure anisotropies $p_x-p_y$ were qualitatively similar to those found from the AdS/CFT framework, but their magnitude grew faster and thus there was never good quantitative agreement. Including the second order correction induces a term that tends to drive the anisotropy to the opposite direction. The net effect is that at very early times there is not even qualitative agreement, but shortly before $t\sim 0.1/\mu$ (which is the estimated upper time limit on the validity of the gradient expansion) the evolution of inhomogeneities reaches a stage where in addition to the qualitative agreement (spatial profiles again having similar shapes) there is even good quantitative agreement, as can be seen in Fig.~\ref{fig:anisotropies2}.
If we were to extend the evolution to later times (say up to $t\sim 0.15/\mu$), then the match with second order viscous hydrodynamics appears to become somewhat less good in the region $x\lesssim 100$, while continuing to improve in the region $x\gtrsim 100$. On the other hand, free streaming appears to remain a very good approximation. Given the very short time window our methods allow us to describe, more work is required to confirm whether the early-time agreement with second order hydrodynamics that we find is accidental or really signals the onset of a hydrodynamic regime.

Our analysis is performed in the infinite coupling limit, so one may ask how finite coupling corrections are expected to affect the results. In particular, for some observables such as the spectral density in photon production in a holographic model of non-equilibrium plasma, it has been reported that finite coupling corrections become significant already at large values of 't Hooft coupling $\lambda$ \cite{Steineder:2013ana,Steineder:2012si}. Other
recent work on finite coupling corrections in the context of thermalization is reported in \cite{Zeng:2013mca,Hassanain:2012uj,Baron:2013cya,Stricker:2013lma}.
However, in the present case the relevant parameters to consider are the shear viscosity $\eta$ and the relaxation time $\tau_{\Pi}$. Finite coupling
corrections to the shear viscosity (more precisely, to its ratio with the entropy density) have been computed in \cite{Buchel:2004di,Benincasa:2005qc,Buchel:2008ac,Buchel:2008sh} and corrections
to the relaxation time have been computed in \cite{Buchel:2008bz}. In both cases, the finite copling corrections become sizable only at small
values of 't Hooft coupling, $\lambda \sim {\cal O}(1-10)$. Based on that, one may expect our results to be applicable at strong but finite coupling.

\subsection{Further examples}\label{sect:examples}

\begin{figure}[h] \begin{center}
\includegraphics[width = 0.46 \textwidth]{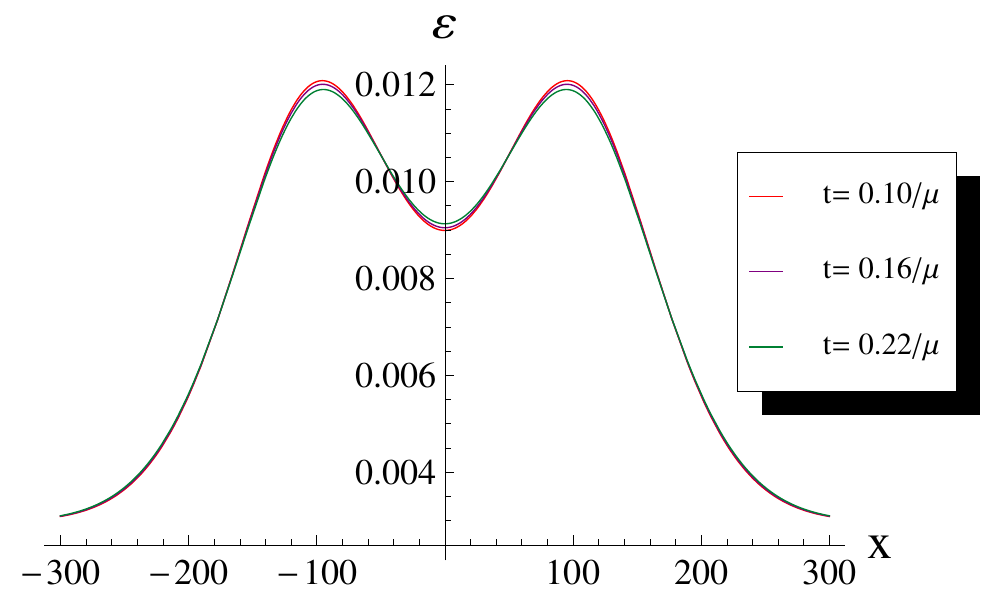}
\hfill
\includegraphics[width = 0.46 \textwidth]{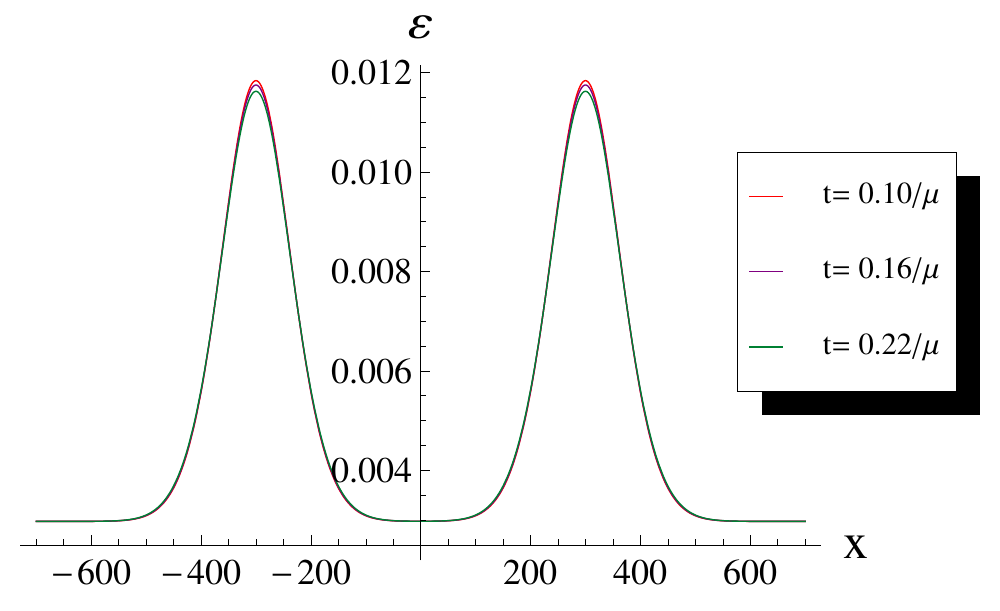} \\
\includegraphics[width = 0.46 \textwidth]{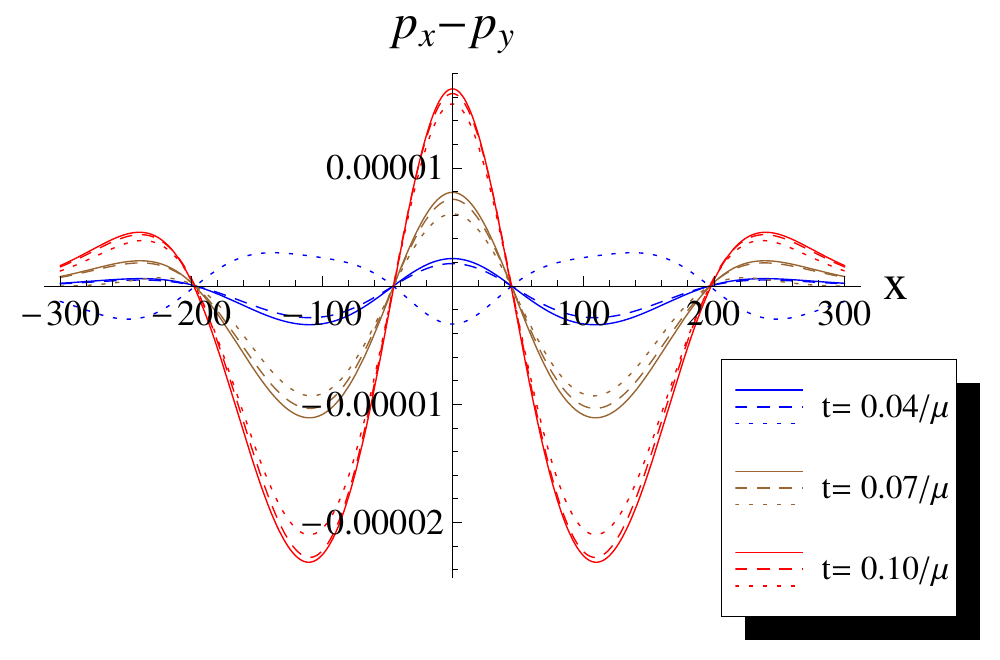}
\hfill
\includegraphics[width = 0.46 \textwidth]{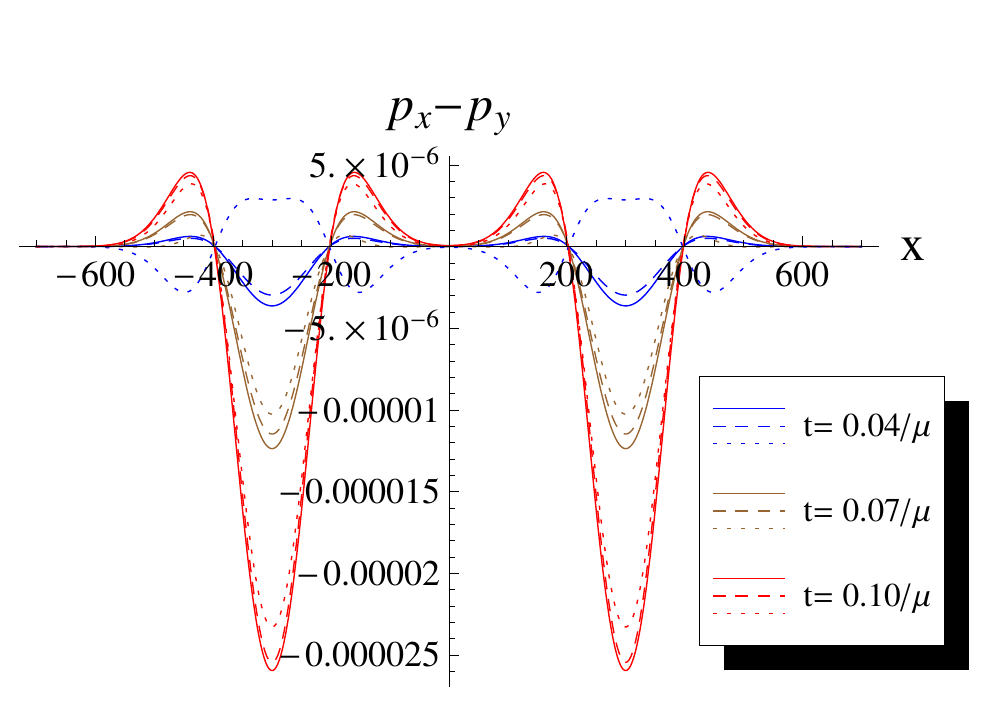}
\caption{Top: the energy density in the rest frame $\varepsilon =p_x +p_y$ (and rescaled with $16 \pi G_N$).  Bottom: the  spatial profile of  the pressures anisotropies $p_x - p_y$ (rescaled by $16 \pi G_N$)  obtained in the AdS/CFT computation (solid lines) compared with those resulting from second order hydrodynamics (dotted) and free streaming (dashed). The source profile has the form \eqref{doublegauss}, with  $\nu =0.5$, $\sigma^2 = 0.1$,  $\mu = 0.01$ and $\eps  = 0.005$. The left column corresponds to $d=2/\mu$, the right one to $d=6/\mu$.} \label{fig:double}
\end{center}
\end{figure}

We have also performed tests with a modified source. Consider first a spatial profile $u(x)$  given by the superposition of two Gaussians with separation $d$ between the center of the two profiles
\be  \label{doublegauss}
\varphi(t, x) = \eps  \( 1+ e^{- \mu^2 \( x - \frac{d}{2} \)^2} +  e^{- \mu^2 \(x+ \frac{d}{2}\)^2}\)  e^{-\frac{\left(t -\nu\right)^2}{\sigma^2}}\, .
\ee
In Fig.~\ref{fig:double} we have plotted the evolution of  the spatial profile of the energy density $\varepsilon$ obtained from the AdS/CFT computation for $d=2/\mu$ (left) and $d=6/\mu$ (right).  The bottom of Fig.~\ref{fig:double} gives a comparison between the evolution of the spatial profile of the pressure anisotropies obtained from AdS/CFT (solid lines), second order hydrodynamics (dotted lines) and free streaming (dashed lines), as in Fig.~\ref{fig:anisotropies2}. It confirms the picture emerging from the single spatial Gaussian analysis.

\begin{figure}[h]
\begin{center}
\includegraphics[width = 0.45 \textwidth]{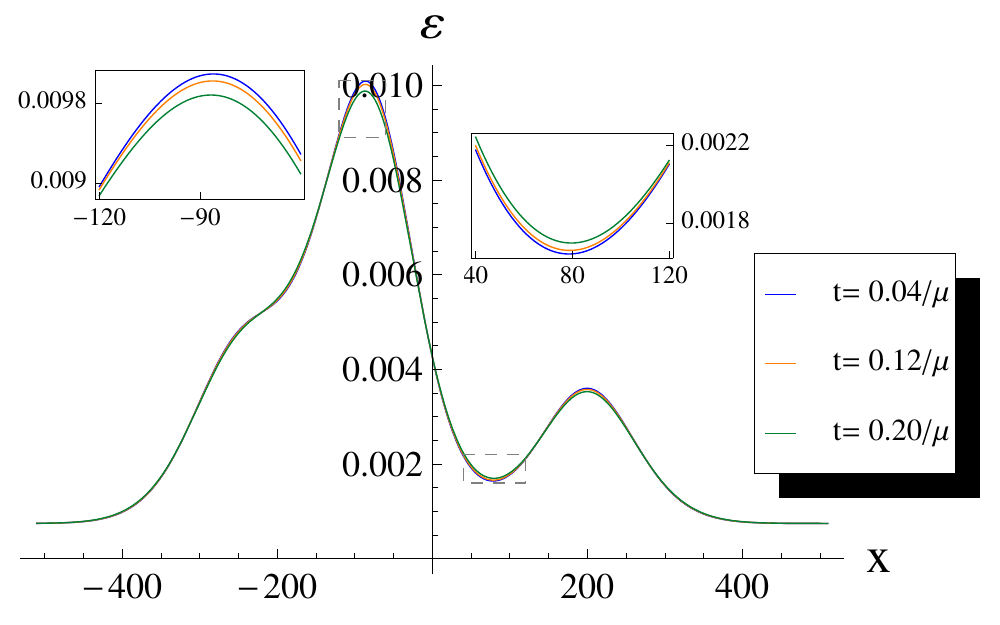}
\hfill
\includegraphics[width = 0.5 \textwidth]{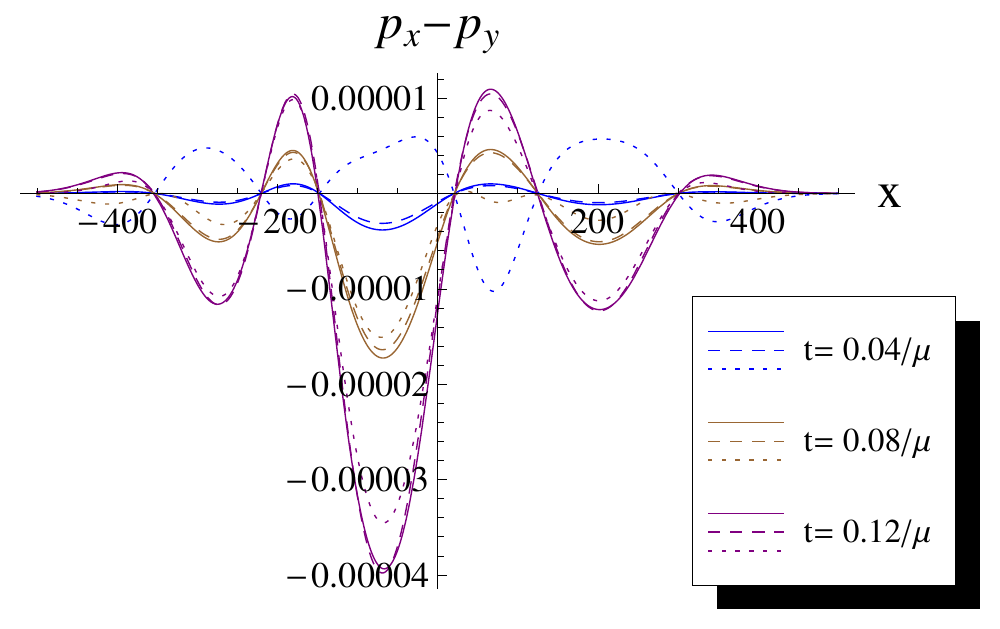}
\caption{Left: the energy density in the rest frame $\varepsilon =p_x +p_y$ (and rescaled with $16 \pi G_N$).  Right: the  spatial profile of  the pressures anisotropies $p_x - p_y$ (rescaled by $16 \pi G_N$)  obtained in the AdS/CFT computation (solid lines) compared with those resulting from second order hydrodynamics (dotted) and free streaming (dashed). The plots are for the source \eqref{source4gauss}, with the same parameter values
$\nu =0.5$, $\sigma^2 = 0.1$,  $\mu = 0.01$ and $\eps  = 0.005$ considered before.}\label{fig:4gauss}
\end{center}
\end{figure}

This conclusion naturally extends to more general superpositions of spatial Gaussians. Let us for instance consider a scalar source with four Gaussian peaks of different size at different locations, on top of a homogeneous background:
\begin{eqnarray}\label{source4gauss}
&& \varphi(t, x) =
\eps \Big( 0.5 + 0.7 e^{- (\mu (x + 250))^2} + 0.5 e^{- (\mu (x + 50))^2}  \\ \nonumber
&& \ \ \ \ \ \ \ \ \ \ \ \ \ \ \ +  0.8 e^{- (\mu (x + 80))^2} + 0.6 e^{- (\mu (x - 200))^2}\Big) e^{-\frac{\left(t -\nu\right)^2}{\sigma^2}}\, .
\end{eqnarray}
The corresponding energy density profile obtained within  the  AdS/CFT approach is plotted in the left panel of Fig.~\ref{fig:4gauss}. In the right panel the pressure anisotropies obtained in the AdS/CFT computation are compared  with those of  free streaming and of  second order viscous hydrodynamics.


\section{Summary and conclusions}\label{sec:summary}

We have generalized the ``naive'' (non-resummed) weak-field perturbation theory part of \cite{Bhattacharyya:2009uu} to include inhomogeneities in a long-wavelength expansion. This has allowed us to study the evolution of the stress tensor of the dual field theory for times short compared to the local inverse ``temperature'', which in turn should be small compared to the local scale of spatial variation. For specific temporal and spatial profiles of the source that injects energy, we have compared the evolution to that of a simple free-streaming model, and have found very good agreement.

In general, hydrodynamics is not meant to describe evolution on (time)scales short compared to the local inverse temperature. However, in various homogeneous models it was found that the stress tensor agreed with that of viscous hydrodynamics much earlier than one had a right to expect \cite{Chesler:2009cy,Beuf:2009cx,Heller:2011ju}. In our inhomogeneous model, for the short time window we are able to explore analytically, we find poor quantitative agreement with first order viscous hydrodynamics. Remarkably, though, second order hydrodynamics does provide very good agreement near the end of the time window in which we trust our computations. This could be a sign of a rapid cross-over from a free-streaming regime to second order hydrodynamics. However, more work is needed to verify whether this is really the beginning of a hydrodynamic regime, rather than an accidental agreement that fails to hold at later times than those we can probe using our methods. To settle this, one would need either an inhomogeneous extension of the resummed perturbation theory of \cite{Bhattacharyya:2009uu}, or a numerical analysis.

In our analysis, we have made several assumptions for technical simplicity: the injected energy only depends on one spatial coordinate, and the field theory lives in a 3d spacetime. In particular, our analysis does not capture the longitudinal expansion of the quark-gluon plasma.  Our results are therefore complementary to those of \cite{Chesler:2009cy,Beuf:2009cx,Heller:2011ju}, where the approach to hydrodynamics is studied for a spatially homogeneous, but longitudinally expanding fluid. Taken together, and with the important caveat mentioned in the previous paragraph, these two results suggest that free streaming followed by second order viscous hydrodynamics may provide a valid description of the space-time evolution of the stress energy tensor in strongly coupled conformal gauge theories over the entire history.

The reader may wonder why free streaming can reproduce the behavior of a strongly coupled gauge theory for any length of time. We do not have a complete answer to this question, but it is tempting to speculate that the short-time behavior of the stress-energy tensor is dominated by the singular behavior of the two-point correlation function of components of the stress-energy tensor. In a conformal theory, the correlation functions have a power-law dependence on the invariant space-time distance and diverge on the light cone. This suggests that free streaming of massless ``particles'' may reproduce the short-time behavior of these correlation functions \cite{Calabrese:2005in}.

We end this paper with a discussion of how our results may help provide a justification for a standard approach to early time evolution of partonic matter in high-energy heavy ion collisions as they are investigated at the CERN Large Hadron Collider (LHC) and the Relativistic Heavy Ion Collider (RHIC).

Due to the large Lorentz $\gamma$ factors of these collisions in the center-of-mass frame, the quark and gluon structure of the colliding nuclei remains essentially frozen during the collision, while the coherence of the nuclear quantum states is broken by soft color exchanges. A large effort was invested into the theoretical investigation of this initial state within a framework called the Color Glass Condensate (CGC) Model. Its central assumption is that nonlinear processes limit the growth of the gluon density at high energy, resulting in its eventual saturation. The model posits that the gluon distribution in the transverse direction attains a universal form -- the color glass condensate -- that is characterized by a single parameter, the saturation scale $Q_s$, which only depends on the size of the nucleus and the beam energy. The very large transverse area density of color charges suppresses all  transverse color correlation beyond a distance  $1/Q_s \ll 1~{\rm fm}/c$ small enough to justify perturbative expansions in the strong coupling constant $\alpha_s(Q_s^2)$. (Note that the running of $\alpha_s$ is actually more involved \cite{Kovchegov:2006vj}, but the basic idea remains the same.) Independently of the specific nature of the CGC model, there exist general arguments that gluon saturation has to occur for sufficiently large collision energies, as described in \cite{Gelis:2012ri}, but the evidence that this regime is already reached in present day experiments is not yet conclusive. One of its consequences, the geometric scaling of cross sections in lepton-hadron interactions, is well established, but geometric scaling seems to be a very general property of high energy interactions and thus does not constitute an unambiguous proof of saturation \cite{Praszalowicz:2012zh}. Keeping these {\em caveats} in mind, it is a well-justified assumption that the initial gluon distribution in high-energy heavy-ion collisions shows saturation. While this saturated state does not necessarily have to be of the universal form of the CGC, it is a reasonable expectation that its properties should be qualitatively similar to those of the CGC. It thus makes sense to use the CGC as a model of the initial state for the study of thermalization in high-energy heavy ion collisions.

Two of the present authors \cite{Muller:2011bb} argued that the fluctuations of the resulting energy density are surprisingly large, suggesting that the time evolution of such fluctuations could be crucial for early thermalization. This calculation was not done strictly within the CGC model,  which assumes Gaussian fluctuations in the transverse color charge distribution. In contrast, the authors of \cite{Muller:2011bb} assumed Gaussian fluctuations in the resulting gauge fields for the purpose of analytical tractability. The CGC makes precise predictions for the transverse correlation function of gauge fields and it is this input which determines the results.
Schenke et al. \cite{Schenke:2012wb} generated fluctuating initial energy densities of colliding nuclei numerically, using the Gaussian color charge fluctuations of the CGC, but did not calculate the resulting energy density correlation function to compare with the analytical model based on Gaussian field fluctuations \cite{Muller:2011bb}.

Some of the many models proposed to describe the surprisingly fast ``hydroization'' are of the type ``Free-streaming plus sudden equilibration'', see
\cite{Broniowski:2008qk} and references given therein. The basic idea of this approach is that after the initial collision/overlap period of the two nucleons,
which for the RHIC energies discussed in \cite{Broniowski:2008qk} is of order 0.1-0.2 fm/c, has defined the initial state for the fireball formation the system is still very far from hydrodynamic behaviour. Basically the kinetic energy completely dominates the dynamics of the partons. As long as this is the case free streaming should provide a good description.  At some time local hydrodynamics takes over. In these models this transition is assumed to be an abrupt one, mainly
for the benefit of computational ease.

Detailed simulations \cite{Schenke:2012hg,Gale:2012rq,Bzdak:2013zma}  use phenomenological models (such as the Color Glass Condensate model) to compute the initial energy deposition and  to evolve it for a short amount of time (of order 0.4 fm/c) using the classical Yang-Mills equations.  These studies give excellent fits to all flow coefficients by assuming that large-amplitude, small-range fluctuations subsequently evolve hydrodynamically. Note that for times short compared to the scale set by the background Yang-Mills field, typically of order 0.2 fm/c, classical Yang-Mills theory is kinetic energy dominated and should therefore be well-approximated by a free-streaming model. Alternate approaches start from fluctuations in nucleon positions and the energy deposition in individual nucleon-nucleon collisions \cite{Alver:2010gr,Petersen:2010cw} and use either free-streaming of particles \cite{Qin:2010pf} or hydrodynamics of  anisotropic fluids \cite{Martinez:2010sc,Ryblewski:2012rr} to bridge the gap to the onset of viscous hydrodynamics.  An important open question, which motivated the present work, is to what extent this ``gluing'' of a phenomenological model describing the initial evolution to viscous hydrodynamics can be justified. If the agreement between free-streaming and second order hydrodynamics that we find near the end of our early-time interval really signals a cross-over (rather than an accidental temporary agreement with hydrodynamics), it provides support for this standard gluing prescription.


\section*{Acknowledgments}

We would like to thank M.~Heller, K.~Rajagopal and especially S.~Minwalla for very useful discussions.

This research is supported by DOE grants  DE-FG02-05ER41367 and DE-FG02-95ER40893, by the Belgian Federal Science Policy Office through the Interuniversity Attraction Poles IAP VI/11 and P7/37, by FWO-Vlaanderen through projects G.0651.11 and G011410N, by the Academy of Finland grants 1127482 and 1268023,  by the National Science Foundation under Grant No.\ PHY11-25915, by the European Science Foundation Holograv Network, by the Finnish Academy of Science and Letters, and by the Vrije Universiteit Brussel through the Strategic Research Program ``High-Energy Physics''.

AB is a Postdoctoral Researcher of FWO-Vlaanderen. FG is Aspirant FWO. VB, JdB, BC, EKV, BM and AS are grateful to the ICTP; VB, AB, BC, FG, EKV and AS to the Lorentz Center; VB and BC to the KITP; BC to the University of Amsterdam and the University of Helsinki; VB, JdB and EKV to the Vrije Universiteit Brussel; BC and EKV to the Newton Institute for hospitality and partial support at various stages of this work.

\setcounter{equation}{0}


\appendix


\section{The bulk solution}
\label{app:bulk}

Here we further comment on and summarize the results obtained in Section \ref{subsec:solution} to leading order in the weak field expansion and up to fourth order in the derivative expansion.

The massless scalar field equation in Poincar\'e AdS${}_4$ in infalling Eddington-Finkelstein
coordinates reads
\be
\partial_r (r^2 \partial_v \phi) +
\partial_v (r^2 \partial_r \phi) +
\partial_r (r^4 \partial_r \phi) +
\partial^2_x \phi + \partial_y^2 \phi=0.
\ee
In Fourier space (for $v,x,y$) this has two solutions, namely
\be \label{eq1}
\phi_{\pm} = e^{ i(\omega\mp\sqrt{\omega^2-k^2})/r} \left( 1 \pm
\frac{  i \sqrt{\omega^2-k^2})}{r} \right) .
\ee
Focusing on the case $\omega^2 < k^2$,  the infalling solution is $\phi_+$.

We can take $k_y=0$, and expand the result in a series around $k_x=0$ to obtain the derivative expansion
in $x$. We find
\be
\phi_+ = 1 + \frac{i\omega}{r} -\frac{k^2}{2 r^2} -\frac{i}{8} \frac{k^4 }{\omega r^3}
-\frac{i k^6}{48\omega^3 r^3} +\frac{ k^6 }{48 \omega^2 r^4} + {\cal O}(k^8).
\ee
We should substitute $i\omega \rightarrow \partial_v $ and $ik \rightarrow \partial_x$
to get the answer in coordinate space.

The result at order $k^4$ therefore becomes
\be
\frac{1}{8} \partial_x^4 \partial_v^{-1} \phi(v,x)  .
\ee
From the structure of (\ref{eq1}), we see that when we expand $\phi_+$ in powers of $k$,
for each power of $k^{2l}$ we get a finite series of terms in $1/r$ that range from $1/r^3$
to $1/r^{l+1}$.

The other solution is $\phi_-$ which is the advanced instead of the retarded solution.
It is not difficult to see that
\be
\phi_-(k,\omega,r) = e^{2i\omega/r} \phi_+(k,-\omega,r)
\ee
Thus, the expansion of $\phi_-$ in powers of $k$ is similar to that of $\phi_+$, except that
there is an overall prefactor of $e^{2i\omega/r}$ which gives rise to an infinite series
in $1/r$. When converting to coordinate space, $i\omega \rightarrow \partial_v$, so that
$e^{2i\omega/r}$ has the effect of shifting $v$ to $v+2/r$.

Overall, at fixed order in spatial derivatives, $\phi_+$ truncates in the $1/r$ expansion and
$\phi_-$ does not. However, we can write $\phi_-$ as a truncated series acting on $\phi(v+2/r,x)$
instead.

Therefore at fourth order in the derivative expansion in $x$ the scalar field solution  consistent with causality (obtained by demanding that the series in $1/r$ should truncate) is then given by
 \be \label{profilo}
 \phi(v,r, x) = \vp(v, x)+ \frac{\dot \vp(v, x)}{r} +  \frac{\vp''(v, x)}{2r^2}  + \frac{\int_{-\infty}^{v} d\tau ~ \vp''''(\tau, x)}{ 8 r^3}\, .
\ee
The solution for the bulk metric is obtained similarly and in Eddington-Finkelstein coordinates  has the form
\bal  \label{metricaEF}
ds^{2} &=   -h(v, r,x)dv^{2} +   2dv \( dr + k(v, r,x) dx\)  +  \\[5 pt]
&\hspace{2.5 cm}+f(v,r,x)^2 e^{B(v,r,x)}dx^2+f(v,r,x)^2 e^{-B(v,r,x)}dy^2  \, ,
\eal
with metric components at fourth order in the derivative expansion
\bea
f(v,r, x) &=& r-\frac{\( \dot\vp(v,x)\)^2}{ 8r} -\frac{\dot\vp(v,x) \vp''(v,x)}{12 r^2} \nonumber \\
&&-\frac{\(\vp''(v,x) \)^2}{48 r^3} -\frac{\dot \vp(v,x)  \int_{- \infty}^{v} d\tau~ \vp''''(\tau,x)}{64 r^3}\,, \nonumber
\eea
\bea  \label{eq:metriccomponents}
h(v,r,x) &=&r^2-\frac{3\(\dot \vp(v,x)\)^2}{4} -\frac{\(\vp'(v, x)\)^{2}}{4} + \frac{C_{2,0}(v,x)}{r}  \nonumber\\
&&+\frac{C_{2,2}(v,x)}{r} + \frac{C_{2,4}(v,x)}{r}  -\frac{\dot \vp(v,x)\dot\vp''(v,x)- 2\ddot \vp(v,x) \vp''(v,x)}{12 r^2}  \nonumber\\
&& -\frac{ K_{2,1}'(v, x) }{12 r^2} -\frac{K_{2,3}'(v,x)}{2r^2}  -\frac{\(\vp''(v,x)\)^2}{6 r^2}    \nonumber \\
&& + \frac{  \dot \vp(v,x)  \int_{- \infty}^{v} d\tau~  \vp''''(\tau,x)  }{16 r^2} -\frac{\beta_{2,2}''(v,x)}{4 r^3 }  + \frac{\(\dot \vp' (v,x)\vp'' (v,x)\)' }{24 r^3}     \nonumber \\
&&- \frac{ \vp''''(v,x) \dot \vp(v,x)   + }{96 r^3} + \frac{  \ddot \vp(v,x)  \int_{- \infty}^{v} d\tau~  \vp''''(t,x)  }{32 r^3}  \,,  \nonumber \\
k(v,r, x) &=& \frac{1}{2}\dot\vp(v,x)\vp'(v,x) + \frac{K_{2,1}(v,x)}{6 r} +  \frac{K_{2,3}(v,x)}{6 r}   \\
&&~~~~~~  - \frac{ \dot\vp'(v,x) \vp''(v,x) }{ 8 r^2 } + \frac{3 \beta_{2,2}(v, x)}{ 4 r^2}  \,,  \nonumber\\
B(v,r,x) &=& \frac{ \( \vp'(v,x)\)^{2} }{4r^2} + \frac{\beta_{2,2}(v, x)}{ r^3} + \frac{ \beta_{2,4,3}(v,x) }{r^3} + \frac{\beta_{2,4,4}(v,x)}{ r^4} \,, \nonumber
\eea
where
\bal
C_{2,0}(v,x) &= \frac{1}{2} \int_{-\infty}^{v} d\tau~\dot\vp(\tau,x)\dddot \vp(\tau,x)\,, \\
C_{2,2}(v,x) &= \frac{1}{4} \int_{-\infty}^{v} d\tau~\left[ -\( \dot\vp' \)^2    -  4 \dot\vp\dot\vp''+ 2 \ddot\vp\vp'' + 2\ddot\vp'\vp' -K_{2,1}'  \right] \,,  \\
K_{2,1}(v,x)
&= 2 \int_{- \infty}^{v} d\tau~\left[  \left(  \int_{- \infty}^{\tau} dt \ddot\vp'(t,x)\ddot\vp(t, x) \right) - \dot\vp(\tau,x)\ddot\vp'(\tau, x) \right]  \\
&= - 2 \int_{- \infty}^{v} d\tau~ \int_{- \infty}^{\tau} dt \dddot\vp'(t,x)\dot\vp(t, x) \,, \\
\beta_{2,2}(v,x)&=\frac{1}{8}\int_{-\infty}^{v} d\tau~ \left[ \(\dot\vp'(\tau,x) \)^2 +\frac{1}{3} K_{2,1}'(\tau, x)\right]  \,,\\
K_{2,3}(v,x) &= \int_{- \infty}^{v} d\tau~ \Big[-2 C_{2,2}'(\tau, x) + 6  \beta_{2,2}'(v,x)  \\
&\hspace{2 cm } +  \dot\vp(\tau,x)\vp'''(\tau, x) -2 \dot\vp'(\tau,x) \vp''(\tau, x)\Big]  \,, \\
C_{2,4}(v,x) &= \frac{1}{4} \int_{-\infty}^{v} d\tau~\Big[ -\( \vp''(\tau,x) \)^2    -   \vp' (\tau,x)\vp''' (\tau,x)  -   K_{2,3}' (\tau,x) \\
&  +   \frac{3}{4} \dot\vp(\tau,x) \int_{-\infty}^{\tau} dw~  [ \vp''''(w,x) ]  \Big] \,,\\
 \beta_{2,4,3}(v,x)&= \frac{1}{8} \int_{-\infty}^{v} d\tau~ \left[ \(\vp'(\tau,x)\vp'''(\tau,x) \) + 2 K_{2,3}'(\tau, x) +  8  \beta_{2,4,4}(\tau,x)\right] \,, \\
 \beta_{2,4,4}(v,x)&= \frac{1}{24} \int_{-\infty}^{v} d\tau~ \left[ \(\dot \vp'(\tau,x)\vp'''(\tau,x) - \dot\vp''(\tau,x) \vp''(\tau,x) \)  + 6  \beta_{2,2}''(\tau,x)\right]\,.
\eal

\setcounter{equation}{0}



\section{From bulk solution to boundary stress-energy tensor}
\label{app:holographicrenormalization}

Given the bulk solution for the metric and for the bulk scalar,  finite physical quantities for the boundary theory can be systematically extracted  through a consistent renormalization of the bulk on-shell action. Here we follow  the procedure of \cite{deHaro:2000xn}.\footnote{Notice however the different conventions used for the Riemann tensor in \cite{deHaro:2000xn}, where it is defined with an overall minus sign with respect to ours.}

The action \eqref{bmact}, in general $d+1$ bulk dimensions, reads
\bal \label{GMGH}
S = \frac{1}{16 \pi G_{N}} \left[  \int d^{d+1}x \sqrt{g}\left(R  - 2 \Lambda-\frac{1}{2}g^{\mu\nu}\del_{\mu}\phi \del_{\nu}\phi \right) - 2 \int d^{d}x \sqrt{\gamma} K  \right]\, ,
\eal
including the boundary  \emph{Gibbons-Hawking} term necessary to have a well defined variational problem with Dirichlet boundary conditions.  $\gamma_{\a\b}$ is  the induced metric at the boundary, $K$ is the trace of the extrinsic curvature of the boundary and $\Lambda = - \frac{d(d-1)}{2}$.

In Fefferman-Graham coordinates the metric and the scalar solutions of the equations of motion following from \eqref{bmact} have the general asymptotic $z \to 0$ structure
\bal \label{feffzz}
\dsty ds^2 &=     R^2_{\ads} \(\frac{dz^{2} }{z^2} + \frac{1}{z^2}g_{\alpha\beta}(x,z) dx^\a dx^\b \) \,  , \\
\dsty g_{\a\b}  &  =  g_{(0),\a\b} + z^2  g_{(2),\a\b} \dots + z^d  g_{(d),\a\b}  +  z^d \log(z) h_{(d),\a\b}+ O(z^{d+1})\, ,
\eal
and
 \be \label{scalarFG}
\phi(x,z) = \phi_{(0)}(x) + z^2 \phi_{(2)}(x) \dots + z^{d }\phi_{(d)}(x) +  z^{d } \log(z) \psi_{(d)} (x)  + O (z^{d+1})  \, ,
\ee
with Dirichlet boundary conditions
\bal \label{diri}
& \dsty \lim_{z \to 0} ~\phi(x,z) = \phi_{(0)}(x)\, ,\\
&\dsty \lim_{z\to 0}~ g_{\a\b}(x,z) = g_{(0),\a\b}(x)\, .
\eal

When evaluated on-shell, the action \eqref{GMGH} diverges. Following \cite{deHaro:2000xn}, it  can be regularized restricting the bulk integral to $z\geq \eps$ and evaluating the boundary term at $z=\eps$. Setting  $R^2_\ads=1$, the regularized action reads
\bea \label{reg}
S_{\eps} &=&\frac{1}{16 \pi G_{N}} \left[ \int_{z \geq  \eps } d^{d+1}x \sqrt{g} \left(R - 2  \Lambda -\frac{1}{2} \(\del \phi \)^2\right) - 2 \int_{z=\eps} d^dx \sqrt{\gamma} K  \right]\,  \nonumber \\
&=&\frac{1}{16 \pi G_{N}}\int d^{d}x  \left[\int_{z\geq \eps} dz \(   - \frac{2 d}{z^{d+1}} \sqrt{\det|g_{\alpha\beta}(x,z)|}  \) \right.  \\
&& \hspace{2cm}\left. \left.- \frac{1}{\eps^{d}} \( -2 d \sqrt{\det|g_{\alpha\beta}(x,z)|}   + 2 z \del_{z} \sqrt{\det|g_{\alpha\beta}(x,z)|}  \)\right|_{z=\eps} \right] \nonumber \, ,
\eea
where we have used the  relation $R -\frac{1}{2} \(\del \phi \)^2 =  2 \frac{d+1}{d-1} \Lambda$ valid on-shell,  $\Lambda = - \frac{d(d-1)}{2}$  and the Fefferman-Graham expansion \eqref{feffzz}.
From this one can extract the IR bulk divergences (terms that diverge when $1/\eps$ is large)  and construct the counterterm action.

We now specialize to  the case  $d=3$ of interest for us  here. Using the  Fefferman-Graham form  of the metric
\be \label{onshell}
 S_{\eps } \simeq   \frac{1}{16 \pi G_{N}}\int d^{d}x \left[ \frac{4}{\eps^{3}} \sqrt{g_0}   - \frac{2}{\eps} \sqrt{g_0}\Tr(g_2) + \dots \right] \, ,
\ee
where dots indicate the finite part of the regularized on-shell action in the limit $\eps \to 0$.

Following  \cite{deHaro:2000xn}, a covariant counterterm action can be obtained in a minimal subtraction scheme writing the coefficients of the divergences  in  terms of the boundary induced metric
\be\label{boundary_induced}
\gamma_{\a\b} =  \frac{1}{ \eps^2}   \left[  g_{(0),\a\b}  +  \eps^2  g_{(2),\a\b} + O(\eps^3) \right] .
\ee

Solving order by order in $z$ the equations of motion following from \eqref{GMGH}, with Dirichlet boundary conditions \eqref{diri}, one finds
\bal \label{consd3}
\phi_{(2)} &= \frac{ \Box \phi_{(0)} }{2}\, ,   \\
g_{(2)\a\b} &= -R_{\a\b} + \frac{1}{2}\del_{\a} \phi_{(0)} \del_\b\phi_{(0)} -\frac{g_{(0)\a\b}}{4} \(-R + \frac{1}{2}  (\del\phi_{(0)})^{2}\)\, ,  \\
\Tr g_{(3)} &= 0 \, ,  \\
\nabla^{\a} g_{(3)\a\b} &=  \phi_{(3)}\del_{\b}\phi_{(0)} \, ,
\eal
where all the contractions and curvature tensors are  obtained using $g_{(0),\a\b}$.
This allows to perturbatively invert the relation \eq{boundary_induced} between $\gamma_{\a\b}$ and ($g_{(0),\a\b}$, $g_{(2),\a\b}$),  and to rewrite the coefficients of the divergences  in \eqref{onshell} in a covariant form. At the order we are interested in
\bal
\frac{\sqrt{g_{0} } }{\eps^3} &= \sqrt{\gamma}\(  1 - \frac{1}{2} \eps^2 \Tr(g_{0}^{-1}g_{(2)})  \)\, ,  \\
\Tr(g_{(2)}) &=-  \frac{\eps}{4} \( R[\gamma] - \frac{1}{2} \( \del \phi \)^2 \)\,  ,
\eal
using which, the covariant  counterterms action in the minimal subtraction scheme  reads
\be
S_{ct} =   \frac{1}{16\pi G_{N}} \int d^{3}x   \sqrt{ \gamma } \left[  -4 -  \( R[\gamma] - \frac{1}{2} \( \del \phi \)^2 \) \right]\, .
\ee
The renormalized action is then defined as
\be
S_{ren} = \lim_{\eps\to 0 } \(S_{\eps} + S_{ct,\eps}\) \, .
\ee

The expectation values of the boundary energy momentum tensor and of the operator $\O$  associated to the massless bulk scalar  are obtained from the on-shell renormalized action as
\be  \label{Td3}
\langle T_{\a\b} \rangle = \frac{2}{\sqrt{|\det g_{(0)} |  }}\frac{\delta S_{\text{ren}}}{\delta g_{(0)}^{\a\b}}  =  \frac{3}{16 \pi G_{N}} g_{(3),\a\b} \, ,
\ee
and
\be   \label{O3}
\langle  \O \rangle = \frac{1}{\sqrt{|\det g_{(0)} |  }}\frac{\delta S_{\text{ren}}}{\delta \phi_{(0)} } = \frac{3}{16 \pi G_{N}} \phi_{(3)}\, .
\ee

\setcounter{equation}{0}

\section{Asymptotic change of coordinates}
\label{app:change}

Here we summarize some of the intermediate results in the change of coordinates from the Eddington-Finkelstein to the Fefferman-Graham form of the metric.

In a compact form,  the metric components of the solution can be written as
\bal
h(v, r,x)&= r^2 \( 1 + \frac{ \H_{(2)} }{r^2} +  \frac{ \H_{(3)} }{r^3} +\frac{ \H_{(4)} }{r^4} ) \)  \\
f(v, r,x)&= r  + \frac{\F_{(1)}}{r} +  \frac{\F_{(2)}}{r^2}+  \frac{\F_{(3)}}{r^3}    \\
k(v, r,x)&= \K_{(0)}+ \frac{\K_{(1)}}{r} \\
B(v, r,x)&= \frac{\B_{(2)}}{r^2} +  \frac{\B_{(3)}}{r^3} +  \frac{\B_{(4)}}{r^4}   \, ,
\eal
while for the scalar field
\be
\phi(v,r,x) = \Psi_{(0)} + \frac{\Psi_{(1)}}{r} + \frac{\Psi_{(2)}}{r^2} + \frac{\Psi_{(3)}}{r^3} \, .
\ee
The explicit expressions of the various coefficients can be identified by comparison with the solution worked out in Section~\ref{subsec:solution}.

Working perturbatively in the radial variable, one can conveniently  write the general form of the  change of coordinates as
\bal \label{cambio}
v(t,\vr,\z ) &=  t + \sum_{n=1}^{\infty} \frac{v_n(t,\z)}{\vr^n}  \\
r(t,\vr, \z) &=  \vr+ \sum_{n=1}^{\infty} \frac{r_n(t,\z)}{\vr^{n-1}} \\
x(t,\vr,\z ) &= \z + \sum_{n=1}^{\infty} \frac{x_n(t,\z)}{\vr^n}  \, .
\eal
The functions $v_n,r_n$ and $x_n$ can be determined by substituting \eqref{cambio} in
\bal
ds^{2} &=   -h(v, r,x)dv^{2} +   2dv \( dr + k(v, r,x) dx\)  +  \\[5 pt]
&\hspace{2.5 cm}+f(v,r,x)^2 e^{B(v,r,x)}dx^2+f(v,r,x)^2 e^{-B(v,r,x)}dy^2  \, ,
\eal
and imposing order by order in $\vr$ that the transformed metric should take the Fefferman-Graham form
\bal
ds^{2} &= \frac{d\vr^{2}}{\vr^2} + \vr^{2}  g_{\a\b}dx^{\a} dx^{\b}  \\
g_{\a\b}& =  g_{(0),\a\b}(t, \chi) + \frac{g_{(2),\a\b}(t,\z)}{\vr^{2}} +  \frac{g_{(3),\a\b}(t,\z) }{\vr^{3}}+ \dots
\eal
In the new coordinates, the scalar field allows the asymptotic expansion
\be
\phi(t,\vr,\z) = \phi_{(0)}(t,\z) +\frac{ \phi_{(2)}(t,\z)}{\vr^{2} } +  \frac{ \phi_{(3)}(t,\z)}{\vr^{3} } + \dots \, .
\ee

From the expression for the stress energy tensor \eqref{eq:vevT} and the operator dual to the bulk scalar field \eqref{eq:vevO}, it  follows that we are interested in the coefficients of metric and of the scalar field only up to $g_{(3),\a\b}$ and $\phi_{(3)}$.

These coefficients can be  fixed by imposing that the transformed metric should take the Fefferman-Graham  form up to corrections of $O(\vr^{-6})$, which is achieved by a coordinate transformation of the form
\bea \label{cambiocord}
v &\to& t -\frac{1}{\vr}  + \frac{v_3}{\vr^3} + O(\vr^{-4}) \, , \nonumber \\
r &\to& \vr \( 1-\frac{\H_{(2)} -3 v_3}{ 3 \vr^2}  + \frac{r_3}{\vr^3}  + O( \vr^{-4} ) \) \, ,\\
x &\to & \z   + \frac{ \K_{(0)} }{ 3\vr^3 } + O( \vr^{-4} )\, ,   \nonumber
\eea
where
\be
v_{3}(t,\z) = \frac{1}{12}\H_{(2)}(t,\z)\,
\ee
and
 \be
 r_{3}(t,\z) = -\frac{1}{6}\( \H_{(3)} + 4 \F_{(2)}- 4 \dot\F_{(1)}-\frac{1}{3}\dot\H_{(2)} + \frac{2}{3}\K'_{(0)}\)\, .
 \ee

The  coefficient  $g_{(0),\a\b}$ is the flat three-dimensional  metric  $g_{(0),\a\b}=\eta_{\a\b} =\text{diag}(-1,1,1)$,
while
\be
g_{(2),\a\b}=
\left(\begin{array}{ccc}
-\frac{1}{2}\H_{(2)} &  \K_{(0)} & 0 \\
 \K_{(0)} & -\frac{1}{2}\H_{(2)} + 2\F_{(1)}+ \B_{(2)}  & 0 \\
0 & 0 & -\frac{1}{2}\H_{(2)} + 2\F_{(1)}- \B_{(2)}
\end{array}\right) \, ,
\ee
and
\be{\footnotesize
g_{(3),\a\b}=\frac{1}{3}\left(
\begin{array}{ccc}
 -2\H_{(3)} &  - \H_{(2)}' - 2 \dot\K_{(0)} + 3 \K_{(1)}  & 0 \\
  - \H_{(2)}' - 2 \dot\K_{(0)} + 3 \K_{(1)}  &- \H_{(3)}  - 3 \B_{(3)}+3 \dot\B_{(2)}  +  \K_{(0)}'
   & 0 \\
 0 & 0  &- \H_{(3)}  + 3 \B_{(3)} -3 \dot\B_{(2)}   -  \K_{(0)}'
\end{array}
\right)\, ,}
\ee
where we have used the relation
\be
 \dot \H_{(2)} + 6 \F_{(2)} - 6 \dot \F_{(1)}  +   \K_{(0)}'   =0 \,
\ee
following from the explicit form of the solution.

For the scalar field, from \eqref{cambiocord}  one obtains
\bal
\phi(t,\vr,\z) &=   \Psi_{(0)}(t,\z) + \frac{\Psi_{(1)}(t,\z) - \dot\Psi_{(0)}(t,\z) }{\vr}  \\
&+ \frac{ \Psi_{(2)}(t,\z)  -\dot\Psi_{(1)}(t,\z) + \frac{1}{2}  \ddot\Psi_{(0)}(t,\z)  }{\vr^2}    \\
&+ \frac{ \Psi_{(3)}(t,\z) -\dot\Psi_{(2)}(t,\z) +  \frac{1}{2} \ddot\Psi_{(1)}(t,\z) -\frac{1}{6} \dddot\Psi_{(0)}(t,\z)    }{ \vr^3} \\
&+  \frac{v_{3}(t,\z) \( \dot \Psi_{(0)}(t,\z)-\Psi_{(1)}(t,\z)  \)  + \frac{1}{3}\(\H_{(2)} \Psi_{(1)}  + \K_{(0)} \Psi'_{(0)} \)}{\vr^3}\\& +O\(\vr^{-4}\) .
\eal
Looking at the explicit form of  the different $\Psi_{(n)}$ one can check that $\phi(t,\vr,\z) $ has exactly the expected form: $\phi_{(1)}$ vanishes and $\phi_{(2)} = \frac{ \Box_{g_{(0)}} \phi_{(0)} }{2}$, as derived in \eqref{consd3}. In fact it is possible to check that all relations in \eqref{consd3} are satisfied.

Substituting the explicit expressions \eqref{phi10}-\eqref{phi14}, we obtain
 \bal
 \phi(t,\vr,\z) & =  \vp(t,\z)- \frac{\vp(t,\z)- \vp''(t,\z) }{2 \vr^2} \\
 &+\frac{ 8 \dddot\vp(t,\z) - 12 \dot \vp''(t,\z) + 3 \int^{t}_{-\infty} d \tau~\vp''''(\tau,\z)  }{24 \vr^3} + O\(\eps^{3},\eta^{5},\vr^{-4}\) \, .
\eal

\setcounter{equation}{0}
\section{The boundary stress-energy tensor}
\label{app:boundary}

For completeness, we here give explicitly the stress-energy tensor components \eqref{stress} up to fourth order in the gradient expansion.

The energy density reads:
\bea \label{Ttt4order}
T_{tt} &=& - \frac{1}{16 \pi G_{N}} \Bigg\{  \frac{1}{2} \int_{-\infty}^{t} d\tau~  \Bigg[ 2 \dot\vp(\tau,x)\dddot \vp(\tau,x)  -\( \dot\vp'(\tau,x)\)^2    \nonumber \\[3 pt]
&&   -  4 \dot\vp(\tau,x)\dot\vp''(\tau,x) + 2 \ddot\vp(\tau,x) \vp''(\tau,x) + 2\ddot\vp'(\tau,x)\vp'(\tau,x)  \nonumber  \\[3 pt]
&& + 2\frac{\del}{\del{x}} \int_{- \infty}^{\tau}ds~\left[ \dot\vp(s,x)\ddot\vp'(s, x) -    \int_{- \infty}^{s} dw~\ddot\vp'(w,x)\ddot\vp(w, x) \right]  \Bigg]  \nonumber \\
&& - \frac 1 2 \int_{-\infty}^t d\tau~ \Big[ (\vp''(\tau,x))^2 +\vp'(\tau,x)\vp'''(\tau,x) \Big] + \frac{3}{8} \int_{-\infty}^t d\tau \dot \vp(\tau,x) \int_{-\infty}^\tau ds~ \vp''''(s,x) \nonumber\\
&&-\frac 1 2 \frac{\del}{\del x} \int_{-\infty}^t d\tau \int_{-\infty}^\tau ds~ \Big[ \dot \vp(s,x)\vp'''(s,x) - 2 \dot \vp'(s,x) \vp''(s,x) \Big]  \nonumber \\
&& +\frac 1 4 \frac{\del^2}{\del x^2} \int_{-\infty}^t d\tau \int_{-\infty}^\tau ds \int_{-\infty}^s d\omega~\left[ - \frac 5 2 (\dot \vp'(\omega,x))^2 - 4 \dot \vp(\omega,x) \dot \vp''(\omega,x) \right. \nonumber\\
&&  \hspace{7 cm } + 2 \ddot \vp(\omega,x) \vp''(\omega,x) + 2 \ddot \vp'(\omega,x)\vp'(\omega,x) \Bigg]  \nonumber\\
&& + \frac 3 4 \frac{\del^3}{\del x^3}   \int_{-\infty}^t d\tau \int_{-\infty}^\tau ds \int_{-\infty}^s d\omega \int_{-\infty}^\omega dp \int_{-\infty}^p dq~\dot \vp(q,x) \dddot \vp'(q,x)\Bigg\} .
\eea
Notice that the term $ \frac{3}{16} \int_{-\infty}^t d\tau \dot \vp(\tau,x) \int_{-\infty}^\tau ds \vp''''(s,x)$ does not give any time dependent contribution after the injection of energy has been completed.

From Eq. \eqref{stress}, we see that $T_{xx}$ and $T_{yy}$, besides by $\H_{(3)} = -8 \pi G_N T_{tt}$, are determined by the combination $- 3 \B_{(3)}  + 3 \dot\B_{(2)}  - \K_{(0)}'$. For $t \ge \delta t$, the latter reduces to $- 3 \B_{(3)}$, since both $\B_{(2)}$ and $\K_{(0)}$ vanish (see Eq. \eqref{expmetric}). Explicitly we have:
\bea
\B_{(3)} &=& \frac 1 8 \int_{-\infty}^t d\tau~ (\dot \vp'(\tau, x))^2 -\frac{ 1}{12} \frac{\del}{\del x} \int_{-\infty}^t d\tau \int_{-\infty}^\tau ds  \int_{-\infty}^s d\omega~ \dddot \vp'(\omega, x) \dot \vp(\omega,x) \nonumber \\
&&+ \frac 1 8 \int_{-\infty}^t d\tau~ \vp'(\tau,x) \vp'''(\tau,x) \nonumber\\
&& + \frac{1}{24}  \int_{-\infty}^t d\tau \int_{-\infty}^\tau ds \Big[ - 5 \dot \vp'(s,x) \vp'''(s,x) - 13 \dot \vp''(s,x) \vp''(s,x) + 6 \dot \vp(s,x) \vp''''(s,x) \Big] \nonumber \\
&& +\frac {1 }{32} \frac{\del^2}{\del x^2}  \int_{-\infty}^t d\tau \int_{-\infty}^\tau ds  \int_{-\infty}^s d\omega \Big[ 11 (\dot \vp'(\omega,x))^2 + 16 \dot \vp(s,x) \dot \vp''(s,x)\nonumber \\
&& \hspace{8 cm }-8 \ddot \vp(s,x) \vp''(s,x) -8 \ddot \vp'(s,x) \vp'(s,x) \Big] \nonumber\\[3pt]
&& -\frac{19}{48} \frac{\del^3}{\del x^3} \int_{-\infty}^t d\tau \int_{-\infty}^\tau ds  \int_{-\infty}^s d\omega \int_{-\infty}^\omega dp \int_{-\infty}^p dq~ \dddot \vp'(q,x) \dot \vp(q,x) \,,
\eea
while the energy flux $T_{tx}$ for $t \ge \delta t $ reduces to
 \bea
 T_{tx} &=& \frac{3}{16 \pi G_N} \K_{(1)} = \frac{3}{16 \pi G_N}  \frac 1 6 \(K_{2,1} + K_{2,3} \) \nonumber \\
 && = \frac{3}{16 \pi G_N} \left\{ -\frac 1 3 \int_{-\infty}^t d\tau \int_{-\infty}^\tau ds ~\dddot \vp'(s,x) \dot \vp(s,x) \right.\nonumber \\
 && +\frac 1 6  \int_{-\infty}^t d\tau \Big[ \dot \vp(\tau,x) \vp'''(\tau,x) - 2 \dot \vp'(s,x) \vp''(s,x) \Big]   \nonumber \\
 && + \frac{1}{24} \frac{\del}{\del x} \int_{-\infty}^t d\tau \int_{-\infty}^\tau ds \Big[ 5(\dot \vp'(s,x))^2 + 8 \dot \vp(s,x) \dot \vp''(s,x) \nonumber \\[3 pt]
 && \hspace{6 cm }- 4 \ddot \vp(s,x) \vp''(s,x) - 4 \ddot \vp'(s,x) \vp'(s,x)\Big] \nonumber \\[3 pt]
&& \left.  -\frac 1 4 \frac{\del^2}{\del x^2} \int_{-\infty}^t d\tau \int_{-\infty}^\tau ds \int_{-\infty}^s d\omega \int_{-\infty}^\omega dp ~\dddot \vp'(p,x) \dot \vp(p,x)\right\}\,.
\eea
Using the factorized ansatz \eqref{eq:factorized}
\be
\vp(t,x) =  u(x) \vp_{0}(t) \, ,
\ee
with
 \be
\begin{aligned}
\left.\begin{array}{ll}
\vp_0(t)=0 \, , &  t \le 0 \\
\vp_0(t)=\eps \tilde \varphi_{0}(t)\, , &  0<t< \delta t  \\
\vp_0(t)=0 \, ,  & t \ge \delta t \,,
\end{array}\right.
\end{aligned}
\ee
the above expressions become
\bea
T_{tt}& =&- \frac{1}{16 \pi G_{N}} \left\{   u(x)^2  \int_{-\infty}^{t} d\tau~ \dot\vp_0(\tau)\dddot \vp_0(\tau)    \right. \nonumber \\
&& - \frac{1}{2} \Big[ u'(x)^2+ 4 u(x)u''(x) \Big]  \int_{-\infty}^{t} d\tau~ \( \dot\vp_0(\tau) \)^2  + \frac{1}{2} \frac{\del^{2}}{\del x^{2}} u(x)^{2} \int_{-\infty}^{t} d\tau~ \ddot\vp_0(\tau) \vp_0(\tau) \nonumber\\
&& +  \frac{1}{2}\frac{\del^{2}}{\del x^{2}} u(x)^{2}\int_{-\infty}^{t} d\tau \int_{-\infty}^{\tau} ds \int_{-\infty}^{s} dw~\dot\vp_0(w)\dddot \vp_0(w) \\
&& + \frac{3}{8} u(x) u''''(x) \int_{-\infty}^t d\tau~ \dot \vp_0(\tau) \int_{-\infty}^\tau ds ~\vp_0(s) \nonumber \\
&&- \frac {1}{8}  \frac{\del}{\del x} \Big[9 u'(x) u''(x) + 6 u(x) u'''(x)\Big] \int_{-\infty}^t d\tau~ \vp_0(\tau)^2  \nonumber \\
&&  +\frac{3}{8} \frac{\del^2}{\del x^2} \Big[ 3(u'(x))^2 + 4 u(x) u''(x) \Big] \int_{-\infty}^t d\tau \int_{-\infty}^\tau ds \int_{-\infty}^s d\omega ~\vp_0(\omega) \ddot \vp_0(\omega) \nonumber \\
&& +\left.\frac{3}{8}  \frac{\del^4}{\del x^4} u(x)^2 \int_{-\infty}^t d\tau \int_{-\infty}^\tau ds \int_{-\infty}^s d\omega \int_{-\infty}^\omega dp \int_{-\infty}^p dq ~ \dot \vp_0(\omega) \dddot \vp_0(\omega) \Bigg]\right\} \,,
\eea
\bea
\B_{(3)} &=& \frac 1 8 (u'(x))^2 \int_{-\infty}^t  d\tau~(\dot \vp_0(\tau))^2  \nonumber\\
&& + \frac{1}{48} \left[ u'(x) u'''(x) -13 (u''(x))^2 + 6 u(x)u''''(x)\right] \int_{-\infty}^t d\tau~\vp_0(\tau)^2 \nonumber\\
&&-\frac{ 1}{24} \frac{\del^2}{\del x^2} u(x)^2 \int_{-\infty}^t  d\tau \int_{-\infty}^\tau  ds \int_{-\infty}^s  d\omega ~\dddot \vp_0(\omega) \dot \vp_0(\omega)\nonumber \\
&& + \frac{1}{32} \frac{\del^2}{\del x^2} (u'(x))^2 \int_{-\infty}^t  d\tau \int_{-\infty}^\tau  ds \int_{-\infty}^s  d\omega \Big[ 11 (\dot \vp_0(\omega))^2 -8 \ddot \vp_0(\omega) \vp_0(\omega) \Big] \nonumber\\
&& + \frac 1 4 \frac{\del^2}{\del x^2} (u(x)u''(x)) \int_{-\infty}^t  d\tau \int_{-\infty}^\tau  ds \int_{-\infty}^s  d\omega \Big[ 2 (\dot \vp_0(\omega))^2 - \ddot \vp_0(\omega) \vp_0(\omega) \Big] \nonumber\\
&& -\frac{19}{96} \frac{\del^4}{\del x^4} u(x)^2   \int_{-\infty}^t d\tau \int_{-\infty}^\tau ds  \int_{-\infty}^s d\omega \int_{-\infty}^\omega dp \int_{-\infty}^p dq ~\dddot \vp_0(q) \dot \vp_0(q)\,, \\[9 pt]
T_{tx} &=&  \frac{3}{16 \pi G_N} \left\{ -\frac 1 6 \frac{\del}{\del x} u(x)^2 \int_{-\infty}^t d\tau \int_{-\infty}^\tau ds ~\dddot \vp_0(s) \dot \vp_0(s) \right. \nonumber \\
&& \left. +\frac 1 6  \Big[ u(x) u'''(x) - 2 u'(x) u''(x)\Big] \int_{-\infty}^t d\tau~  \dot \vp_0(\tau) \vp_0(\tau)    \right. \nonumber \\
&& + \frac{1}{24} \frac{\del}{\del x} (u'(x))^2 \int_{-\infty}^t d\tau \int_{-\infty}^\tau ds \Big[ 5(\dot \vp_0(s))^2 - 4 \ddot \vp_0(s) \vp_0(s) \Big] \nonumber \\
&& + \frac{1}{6} \frac{\del}{\del x} (u(x) u''(x)) \int_{-\infty}^t d\tau \int_{-\infty}^\tau ds \Big[ 2(\dot \vp_0(s))^2 -  \ddot \vp_0(s) \vp_0(s) \Big] \nonumber \\
&& \left.  -\frac 1 8 \frac{\del^3}{\del x^3} u(x)^2  \int_{-\infty}^t d\tau \int_{-\infty}^\tau ds \int_{-\infty}^s d\omega \int_{-\infty}^\omega dp~ \dddot \vp_0(p) \dot \vp_0(p)\right\} \,.
\eea

\setcounter{equation}{0}

\setcounter{equation}{0}



\end{document}